\documentclass[showpacs,twocolumn,showkeys,amsmath,amssymb,pra]{revtex4-1}

\usepackage{color}
\usepackage{graphicx}   
\newcommand{\rd}{{\mathrm d}}
\newcommand{\re}{{\mathrm e}}
\newcommand{\ri}{{\mathrm i}}                   

\begin{document}

\title[Unconventional quasienergy bands]
	{Escape of ultracold atoms from unconventional quasienergy bands}

\author{Onno R. Diermann}
\author{Martin Holthaus}

\affiliation{Institut f\"ur Physik, Carl von Ossietzky Universit\"at, 
	D-26111 Oldenburg, Germany}
                  
\date{October 30, 2026}

\begin{abstract}
Employing the Floquet picture, we study the dynamics of ultracold 
atoms in one-dimensional optical lattices that are subjected to both a static 
bias and resonant time-periodic forcing. This combination leads to the emergence 
of quasienergy bands that differ markedly from the usual energy bands encountered 
with one-dimensional lattice potentials, being perforated by a plethora of avoided 
crossings with almost-free particle states, thereby encoding the characteristic 
escape times. Our analysis reveals a far-reaching analogy between the 
loss of ultracold atoms from periodically driven optical lattices 
and the quantum decay problem. We also predict that decaying Wannier-Stark 
ladder states can be stabilized through the action of a strong time-periodic force.
\end{abstract} 


\maketitle 


\section{Introduction}
\label{sec:1}

Within the last two decades, the experimental and theoretical study of 
ultracold atoms in optical lattices has matured into a most prolific subfield 
of physics~\cite{GreinerEtAl02,LewensteinEtAl12}. In particular, a steadfastly 
increasing amount of work is presently being done on the dynamics of ultracold 
atoms in time-periodically driven optical lattices~\cite{Eckardt17}, largely 
due to the recognition that one may utilize time-periodic forcing for 
systematically equipping a quantum system with properties that it does not 
have in the absence of the force~\cite{GoldmanDalibard14,Holthaus16}. 

These developments necessarily have led to the question how lattice atoms heat 
up in response to an externally imposed time-periodic drive~\cite{PichlerEtAl13,
WeinbergEtAl15,StraeterEckardt16,ReitterEtAl17}. Phrased more generally, the 
overarching question how energy absorption proceeds in periodically driven 
quantum systems is now under active investigation~\cite{AbaninEtAl15,
MoriEtAl16,BukovEtAl16,AbaninEtAl17,HoEtAl18,IshiiEtAl18}.  

With the present paper we wish to draw attention to a particular
scenario that can be realized with optical lattices that are both tilted 
by a static force and driven by a time-periodic force in a resonant manner. 
As will be shown, this set-up leads to the emergence of unconventional 
quasienergy bands that do not conform at all to the universal description 
achieved by Kohn for the smooth energy bands hosted by one-dimensional crystal 
lattices~\cite{Kohn59}, but are strongly lacerated, due to the omnipresent 
coupling of tightly bound lattice Floquet states to continuum-like above-barrier 
states. This results in the excitation of cold atoms initially stored 
in the lattice, and in their eventual escape. Such coupling between two fairly 
different types of states places the loss of ultracold atoms from 
these unconventional quasienergy bands in close conceptual proximity to the 
barrier penetration problem as encountered, for instance, in the theory of 
the $\alpha$-decay, with the characteristic escape time scale being 
encoded in the spectral properties of the initial quasienergy band. 

We proceed as follows: In Sec.~\ref{sec:2} we provide a general explanation 
for the emergence of unconventional quasienergy bands, and review in 
Sec.~\ref{sec:3} their tight-binding description~\cite{Holthaus92,Zak93}.
The corresponding single-band Floquet states approximate the initial
quasienergy states for the excitation process, whereas the final 
ones are constructed in Appendix~\ref{app:1}. In Sec.~\ref{sec:4} we then 
present the results of full-fledged, approximation-free numerical calculations 
for one-dimensional, tilted and resonantly driven optical lattices, showing 
striking examples for the ``ragged'' quasienergy dispersion relations of 
unconventional bands. Appendix~\ref{app:2} provides background information 
on the so-called box stabilization method that allows one, for instance, to 
compute the decay times for typical barrier penetration problems; this 
technique is adapted in Sec.~\ref{sec:4} for estimating escape times for 
ultracold atoms in unconventional quasienergy bands. Some conclusions 
concerning future experiments are drawn in the final Sec.~\ref{sec:5}, 
and parallels are sketched to earlier experimental works in 
other fields.

\section{Quasienergy bands for tilted lattices}
\label{sec:2}

We consider a particle of mass~$M$ moving on a one-dimensional lattice 
specified by a potential $V(x)$ while being acted on by a spatially
homogeneous force $F(t)$, as described by the Hamiltonian   
\begin{equation}
	H(t) = \frac{p^2}{2M} + V(x) - x F(t) \; .
\label{eq:OHA}
\end{equation}
We denote the lattice constant by $a$, so that 
\begin{equation}
	V(x) = V(x+a) \; .
\label{eq:LAP}
\end{equation}	
In addition, we require the force to be periodic in time with period~$T$,
\begin{equation}
	F(t) = F(t+T) \; .
\label{eq:DRF}
\end{equation}
Since then also $H(t) = H(t+T)$, this implies that the time-dependent
Schr\"odinger equation 
\begin{equation}
	\ri \hbar \frac{\partial}{\partial t} \psi(x,t)
	= H(t) \, \psi(x,t)
\label{eq:TDS}
\end{equation}
admits a complete set of Floquet states~\cite{Holthaus16,Sambe73,
FainshteinEtAl78}, that is, of solutions having the particular form
\begin{equation}
	\psi_{\varepsilon}(x,t) = 
	u_{\varepsilon}(x,t) \exp(-\ri\varepsilon t/\hbar)
\label{eq:FLS}
\end{equation}
with $T$-periodic Floquet functions
\begin{equation}
	u_{\varepsilon}(x,t) = u_{\varepsilon}(x,t+T) \; .
\label{eq:FFU}	
\end{equation}
Inserting these Floquet states~(\ref{eq:FLS}) into the Schr\"odinger 
equation~(\ref{eq:TDS}), it follows that the Floquet functions~(\ref{eq:FFU})
obey the eigenvalue equation
\begin{equation}
	K u_{\varepsilon}(x,t) = \varepsilon \, u_{\varepsilon}(x,t)			 
\label{eq:QEP}
\end{equation}
for the quasienergies $\varepsilon$, where 
\begin{equation}
	K = H(t) + \frac{\hbar}{\ri} \frac{\partial}{\partial t} 
\end{equation}
is the quasienergy operator. Note that this eigenvalue problem is posed on a 
so-called extended Hilbert space $\mathcal H$ of $T$-periodic functions, for 
which the time~$t$ plays the role of an additional coordinate~\cite{Sambe73,
FainshteinEtAl78,Howland79}.

In order to exploit, besides the temporal translation symmetry of the driving 
force~(\ref{eq:DRF}) incorporated into the Floquet states~(\ref{eq:FLS}), also 
the spatial translation symmetry of the lattice potential~(\ref{eq:LAP}), we 
follow a line of reasoning laid out by Zak~\cite{Zak93} and consider ``electric
translations'' by the lattice period~$a$, as mediated by the operators
\begin{equation}
	\beta(a) = \exp\left(\frac{\ri}{\hbar}[p + A(t)]a \right) \; ,
\label{eq:ELT}
\end{equation}
where the force's potential $A(t)$ has been introduced through the condition
\begin{equation}
	-\frac{\rd }{\rd t} A(t) = F(t) \; ; 
\label{eq:DFA}
\end{equation}
observe that this definition fixes $A(t)$ only up to a yet arbitrary constant
of integration, $\hbar k_0$. Now there is an apparent mathematical detail with 
profound implication for physics: Suppose that the force~$F(t)$ contains a dc 
component, such as 
\begin{equation}
	F(t) = F_0 + F_1\cos(\omega t + \varphi_0) 
\label{eq:ADF}	
\end{equation}	
with $T = 2\pi/\omega$. Then this dc force $F_0$ gives rise to a secular 
contribution to the potential $A(t)$ that grows linearly in time,
\begin{equation}
	A(t) = \hbar k_0 
	-F_0 t - \frac{F_1}{\omega}\sin(\omega t + \varphi_0) \; ,
\label{eq:POT}
\end{equation}	
so that $A(t)$ is {\em not\/} $T$-periodic. As a consequence, the 
operators~(\ref{eq:ELT}) effectuating the electric translations do not act 
on the extended Hilbert space~${\mathcal H}$ of $T$-periodic functions, 
{\em unless\/} one has an equality $F_0 T a/\hbar = r \times 2\pi$ with 
integer $r = 0,\pm 1, \pm 2, \ldots\,$, so that the strength $F_0$ of the 
dc~force is resonantly related to the driving frequency~$\omega = 2\pi/T$: 
\begin{equation}	
	F_0 a = r \times \hbar\omega \; .
\label{eq:RES}
\end{equation}	
In these particular cases~(\ref{eq:RES}) it is meaningful to consider the 
commutator of the quasienergy operator~$K$ and the electric 
translations~$\beta(a)$, yielding 
\begin{equation}
	[ K, \beta(a) ] = 0 
	\quad \text{on } {\mathcal H}  \; ,
\end{equation}	
which ensures that the $T$-periodic quasienergy eigenfunctions of $K$ can be 
chosen such that they simultaneously are eigenfunctions of $\beta(a)$. Since 
one has $\beta^2(a) = \beta(2a)$, we write the eigenvalues of the unitary
operator $\beta(a)$ as $\re^{\ri k a}$, obtaining
\begin{eqnarray}
	\beta(a) \, u_{\varepsilon}(x,t) 
	& = & u_{\varepsilon}(x+a,t) \re^{\ri A(t) a/\hbar}
\nonumber \\	& = &	
	\re^{\ri k a} u_{\varepsilon}(x,t) \; .
\label{eq:BAU}
\end{eqnarray}		
This creates a link between the respective quasienergy~$\varepsilon$ and
the corresponding wave number~$k$, giving rise to quasienergy-quasimomentum 
dispersion relations~$\varepsilon(k)$. Invoking an index~$n$ to distinguish 
different quasienergy bands, we thus have proven the following 
assertion~\cite{Zak93}:

Provided the resonance condition~(\ref{eq:RES}) is satisfied, the 
time-dependent Schr\"odinger equation~(\ref{eq:TDS}) with the 
Hamiltonian~(\ref{eq:OHA}) admits spatio-temporal Bloch waves 
\begin{equation}
	\psi_{n,k}(x,t) = u_{n,k}(x,t) 
	\exp\!\big(-\ri \varepsilon_n(k) t/\hbar\big)
\label{eq:STB}
\end{equation}
with
\begin{equation}
	u_{n,k}(x+a,t) = \re^{\ri[k - A(t)/\hbar]a} u_{n,k}(x,t)
\label{eq:SBC}
\end{equation}
and
\begin{equation}
	u_{n,k}(x,t+T) = u_{n,k}(x,t) \; ;
\label{eq:TBC}	
\end{equation}	
in the absence of a dc force, that is, for $r = 0$, these solutions reduce
to the conventional Floquet-Bloch states~\cite{Holthaus16}. Here we require
$r \ne 0$, and designate bands with $r \ne 0$ as {\em unconventional 
quasienergy bands\/}.

Because the complex exponential function is $2\pi$-periodic, the eigenvalues
$\re^{\ri k a}$ appearing in the second line of Eq.~(\ref{eq:BAU}) are defined
only up to an integer multiple of the reciprocal lattice constant $2\pi/a$,
implying the customary identification 
\begin{equation}
	k \equiv k + \frac{2\pi}{a}
\end{equation}
which in solid-state physics motivates the construction of Brillouin zones of 
the reciprocal lattice~\cite{AshcroftMermin76}. By the same token, the Floquet
multipliers $\exp(-\ri\varepsilon_n(k) T/\hbar)$ constitute the eigenvalues of 
the time evolution operator~$U(t+T,t)$ over one driving cycle,
\begin{equation}
	\psi_{n,k}(x,t+T) = 
	\exp\!\big(-\ri\varepsilon_n(k) T/\hbar\big) \, \psi_{n,k}(x,t) \; ,
\end{equation}
so that the quasienergies are determined only up to an integer multiple of
$2\pi \hbar/T = \hbar\omega$:
\begin{equation}
	\varepsilon_n(k) \equiv \varepsilon_n(k) + \hbar\omega \; .
\end{equation}
This identification establishes a Brillouin zone scheme for quasienergies, 
with $\hbar\omega$ denoting the width of the quasienergy Brillouin zone. In 
many cases it is helpful to consider the analog of the {\em extended\/} 
Brillouin zone scheme known from solid state physics~\cite{AshcroftMermin76} 
by introducing sets of equivalent quasienergy representatives    
\begin{equation}
	\varepsilon_{n,m}(k) = \varepsilon_n(k) + m\hbar\omega
\label{eq:REP}	
\end{equation}	 
equally spaced by the quasienergy Brillouin zone width $\hbar\omega$, starting 
from an arbitrary representative $\varepsilon_n(k) = \varepsilon_{n,0}(k)$, 
and employing the integer~$m = 0, \pm 1, \pm 2, \ldots$ to designate its 
further satellites; the spatio-temporal Bloch waves~(\ref{eq:STB}) then are 
written as
\begin{equation}
	\psi_{n,k}(x,t) = u_{n,k,m}(x,t) 
	\exp\!\big(-\ri \varepsilon_{n,m}(k) t/\hbar\big)
\end{equation}
with properly $T$-periodic eigenfunctions
\begin{equation}
	u_{n,k,m}(x,t) = u_{n,k}(x,t) \re^{\ri m \omega t} \; .	
\label{eq:FUM}
\end{equation}

\begin{figure}[t]
\begin{center}
\includegraphics[width = 0.9\linewidth]{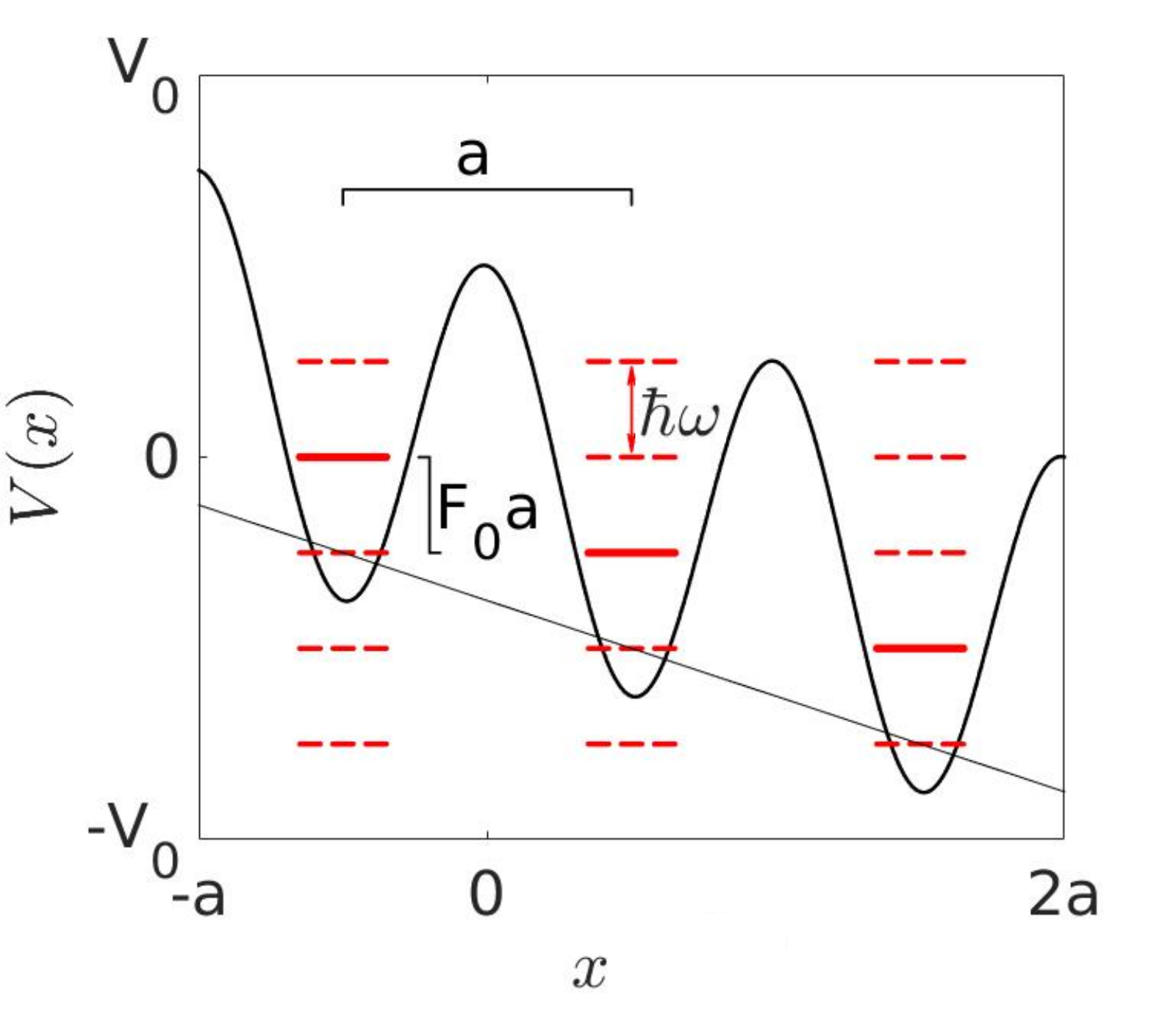}
\end{center}
\caption{(Color online) Sketch of an optical cosine lattice~(\ref{eq:OCL})
	with lattice constant~$a = \pi/k_{\rm L}$ that is tilted by a linear 
	potential~$-F_0x$. Heavy full lines symbolize local energy levels, 
	forming a Wannier-Stark ladder with rungs spaced by $F_0a$. If this 
	spacing is precisely compensated by the energy of $r$~``photons'' 
	$\hbar\omega$, as indicated here for $r = 1$ by broken thinner lines, 
	the resulting {\em quasienergetic alignment\/} permits the existence 
	of uniformly extended spatio-temporal Bloch waves~(\ref{eq:STB}) 
	despite the tilt.}      
\label{F_1}
\end{figure}

This ``extended'' viewpoint leads to an intuitive interpretation of the
unconventional quasienergy bands, depicted in Fig.~\ref{F_1}. According to 
elementary band structure theory~\cite{AshcroftMermin76,Kohn59}, the emergence 
of a customary Bloch energy band for the lattice potential $V(x)$ may be 
understood as being due to the hybridization of energetically aligned Wannier 
states $w_\ell(x)$ localized at the individual lattice sites labeled by~$\ell$.
The addition of the static tilt $-F_0x$ then lifts the energetic alignment, 
forbidding Bloch band formation and giving rise to the so-called Wannier-Stark 
ladder of localized metastable states with resonance energies 
$E_\ell = \ell F_0 a$~\cite{Wannier60,Avron82,GlueckEtAl02}. The further 
addition of a dc-free time-periodic force with an angular frequency~$\omega$ 
obeying the above condition~(\ref{eq:RES}) implies that the spacing $F_0a$ 
between the rungs of the Wannier-Stark ladder is precisely compensated by the 
energy of $r$~``photons''. From the perspective of the extended Brillouin zone 
scheme, which endows each local energy level $\ell F_0 a$ with an infinite 
number of satellites spaced by $\hbar\omega$, this means that the tilted local 
energy levels are quasienergetically realigned, as sketched in Fig.~\ref{F_1} 
for $r = 1$. It is this quasienergetic alignment that underlies the formation 
of quasienergy bands~\cite{Zak93,HolthausEtAl95,DreseHolthaus96}, just as 
energetic alignment underlies energy band formation.

These formal considerations now need to be substantiated by computations of
dispersion relations $\varepsilon_n(k)$ under realistic conditions, prompting
for solutions of the quasienergy eigenvalue problem~(\ref{eq:QEP}) with the 
combined spatio-temporal boundary conditions~(\ref{eq:SBC}) and (\ref{eq:TBC}).   
For numerical purposes the spatial boundary condition~(\ref{eq:SBC}) is somewhat
cumbersome; it is often more convenient to impose periodic boundary conditions 
that allow one to expand the wave functions with respect to a basis set of 
time-independent spatially periodic functions. Therefore, introducing the 
operators
\begin{equation}
	M_k(t) = \exp\!\big( -\ri [k - A(t)/\hbar] x \big) \; ,
\label{eq:GMO}
\end{equation}  
we perform the transformations 
\begin{equation}
	\widetilde{\psi}_k(x,t) = M_k(t) \, \psi(x,t)
\label{eq:GMT}
\end{equation}
that bring the original Schr\"odinger equation~(\ref{eq:TDS}) into the 
spatially periodic form
\begin{equation} 	
	\left[ \frac{1}{2M} \Big( p + \hbar k - A(t) \Big)^2 + V(x) 
	- \ri\hbar \frac{\partial}{\partial t} \right] 
	\widetilde{\psi}_k(x,t) = 0 \; .
\label{eq:TRE}
\end{equation}
Thus, these transformations~(\ref{eq:GMT}) are closely related to the familiar 
G\"oppert-Mayer transformation~\cite{GoeppertMayer31}, converting the 
Hamiltonian from the ``length gauge'' employed in the formulation~(\ref{eq:OHA})
to the ``velocity gauge'' showing up here. Observe that the 
operators~(\ref{eq:GMO}) generally are {\em not\/} periodic in time, even if 
the resonance condition~(\ref{eq:RES}) is satisfied. Writing
\begin{equation}
	\widetilde{u}_{n,k}(x,t) = M_k(t) u_{n,k}(x,t) \; ,
\end{equation}
we therefore have
\begin{equation}
	\widetilde{u}_{n,k}(x+a,t) = \widetilde{u}_{n,k}(x,t) 		 
\label{eq:PBS}
\end{equation}
as desired, whereas $T$-periodicity is lost:
\begin{equation}
	\widetilde{u}_{n,k}(x,t+T) = \widetilde{u}_{n,k}(x,t) 
	\re^{-\ri 2\pi r x/a}\; .	
\end{equation}
Denoting the time evolution operator pertaining to the transformed 
equation~(\ref{eq:TRE}) by $\widetilde{U}_k(t,0)$, the Floquet multi\-pliers
$\exp(-\ri\varepsilon_n(k) T/\hbar)$ then are obtained as eigenvalues of the
composite operator
\begin{equation}
	U_k(T,0) = M_k^\dagger(T) \, \widetilde{U}_k(T,0) \, M_k(0) \; .
\label{eq:STR}
\end{equation}

\section{The single-band tight-binding model}
\label{sec:3}

The previous deliberations have skipped over a technical detail: As noted, the 
defining requirement~(\ref{eq:DFA}) determines the potential $A(t)$ only up to 
an arbitrary constant of integration, $\hbar k_0$. Evidently that constant 
would lead to a likewise arbitrary shift of the wave number~$k$ in both
the transforming operators~(\ref{eq:GMO}) and the transformed Schr\"odinger
equations~(\ref{eq:TRE}), which means that the quasienergy  dispersion 
relations $\varepsilon_n(k)$ are still not specified uniquely. Any shifted 
assignment $\varepsilon_n(k-k_0)$ would do as well, unless additional 
requirements have to be fulfilled, as corresponding to a particular gauge. 
In order to illustrate this point, and to prepare the subsequent discussion 
of the eventual escape of ultracold atoms from 
periodically driven and tilted optical lattices, we now consider a single-band 
tight-binding approach, as is often considered appropriate for sufficiently 
deep lattices. Neglecting all higher bands, let 
$w_\ell(x) \equiv \langle x | \ell \rangle$ denote the Wannier function 
belonging to the lowest energy band that is located at the $\ell$-th lattice 
site. Accounting for nearest-neigbor tunneling only, and resorting to bra-ket 
notation, the infinite-lattice Hamiltonian then takes the paradigmatically 
simple form  
\begin{equation}
	H_0 = -\frac{W}{4} \sum_\ell 
	\Big( | \ell + 1 \rangle \langle \ell | 
	    + | \ell \rangle \langle \ell + 1 | \Big) \; ;
\end{equation}
Wannier functions $\langle x | \ell \rangle$ and hopping matrix 
elements pertaining to optical cosine lattices may be found in 
Refs.~\cite{BoersEtAl07,EckardtEtAl09}. Again denoting the lattice period
by~$a$, the Bloch waves
\begin{equation}
	| \psi_k \rangle = \sum_\ell \exp(\ri k \ell a) \, | \ell \rangle 
\end{equation}		    
solve the stationary Schr\"odinger equation 
$H_0 | \psi_k \rangle = E(k) | \psi_k \rangle$ and thereby constitute a 
cosine energy band of width~$W$, 
\begin{equation}
	E(k) = -\frac{W}{2} \cos(ka) \; .
\end{equation}	
Next, employing the appropriately discretized position operator 
\begin{equation}
	x = \sum_\ell |\ell\rangle a \ell \langle\ell| \; ,
\label{eq:POS}
\end{equation}	  
we add time-periodic forcing
\begin{equation}
	H_{\rm int}(t) = -F(t)a  \sum_\ell |\ell\rangle \ell \langle\ell|
	\; , 
\end{equation}
obtaining the total Hamiltonian $H(t) = H_0 + H_{\rm int}(t)$ in the temporally
periodic guise~(\ref{eq:OHA}). The transformations~(\ref{eq:GMT}), as now 
implemented through the prescription 
\begin{equation}
	| \widetilde{\psi}_k(t) \rangle = \exp\!\left( -\ri [k -A(t)/\hbar] a 
	\sum_\ell |\ell\rangle \ell \langle\ell| \right) | \psi(t) \rangle
	\; ,
\label{eq:DTR}
\end{equation}	 
then produce the Schr\"odinger equations
\begin{equation}
	\left[ \widetilde{H}_k(t) - \ri\hbar\frac{\rd}{\rd t} \right]
	| \widetilde{\psi}_k(t) \rangle = 0 \; ,
\label{eq:TSE}
\end{equation}
where the new Hamiltonian operators $\widetilde{H}_k(t)$ feature Peierls-like 
hopping phases as known from the theory of Bloch electrons in magnetic 
fields~\cite{Peierls33,Kohn59b}, 
\begin{equation}
	\widetilde{H}_k(t) =  -\frac{W}{4} \sum_\ell 
	\Big( \re^{-\ri[k - A(t)/\hbar]a}| \ell + 1 \rangle \langle \ell | 
        + \text{h.c.} \Big) \; ,
\label{eq:HPP}	 	  
\end{equation}	
with ``h.c.'' indicating the Hermitian conjugate of the preceding term.
Brief inspection reveals that these Schr\"odinger equations~(\ref{eq:TSE}) 
are solved by the functions
\begin{equation}  
  	| \widetilde{\psi}_k(t) = \sum_\ell | \ell \rangle
	\exp\left( -\frac{\ri}{\hbar} \int_{t_0}^t \! \rd \tau \,
	E\big(k - A(\tau)/\hbar\big) \right) \; ,
\label{eq:AFS}
\end{equation}	
where $t_0$ is an arbitrary initial time. If we now impose ac-dc forcing of the
type~(\ref{eq:ADF}), making sure that the resonance condition~(\ref{eq:RES}) 
is satisfied, the Hamiltonians~(\ref{eq:HPP}) not only are periodic in space, 
as ensured by the transformations~(\ref{eq:DTR}) anyway, but simultaneously 
periodic in time. This is due to the fact that the Wannier 
states~$|\ell\rangle$ constitute eigenstates of the position 
operator~(\ref{eq:POS}) with eigenvalues~$\ell a$ that are integer multiples 
of the lattice period~$a$, so that the transformations~(\ref{eq:DTR}) are 
$T$-periodic, unlike their more general continuous counterparts~(\ref{eq:GMT}). 
As a consequence, these wave functions~(\ref{eq:AFS}) actually are Floquet 
states: The relation~(\ref{eq:RES}) guarantees that the argument 
$k - A(t)/\hbar$ traverses $r$~quasi\-momentum Brillouin zones during 
one driving cycle~$T$, since $F_0 T = r \times 2\pi\hbar/a$, so that the 
functions $f_k(t) = E\big(k - A(t)/\hbar\big)$ are $T$-periodic. Hence, their 
integrals determining the time-dependence of the solutions~(\ref{eq:AFS}) 
separate into one contribution that again is $T$-periodic in time, stemming 
from the oscillating components of $f_k(t)$, and a secular contribution 
stemming from their respective $k$-dependent average that grows linearly in 
time, thereby establishing the quasienergy band  
\begin{equation}
	\varepsilon(k) = \frac{1}{T} \int_0^T \! \rd \tau \, 
	E\big(k - A(\tau)/\hbar\big)
	\quad \bmod \hbar\omega \; . 
\label{eq:QEB}
\end{equation}  
Inserting the potential~(\ref{eq:POT}), and utilizing the Jacobi-Anger 
relation
\begin{equation}
	\re^{\ri z \sin \alpha} = \sum_{\nu = -\infty}^\infty 
	\re^{\ri \nu \alpha} J_\nu(z) \; ,
\end{equation}
where $J_\nu(z)$ denotes the Bessel function of integer 
order~$\nu$~\cite{AbramowitzStegun72},
Eq.~(\ref{eq:QEB}) yields the explicit expression
\begin{equation}
	\varepsilon(k) = 
	(-1)^{r+1} \frac{W}{2} J_r\!\left(\frac{F_1a}{\hbar\omega}\right)
	\cos\!\Big( (k-k_0)a - r\varphi_0\Big) 
	\; \bmod \hbar\omega \; . 
\label{eq:QTB}	
\end{equation}
This agrees with the result stated by Zak~\cite{Zak93}, additionally revealing
the freedom of choice for~$k_0$. Here we suggest a particular gauge: For {\em 
conventional\/} quasienergy bands emerging in the absence of a tilt~$F_0$, 
so that $r = 0$ in the resonance condition~(\ref{eq:RES}), it seems reasonable 
to require that each individual quasienergy $\varepsilon(k)$ reduce to the 
energy $E(k)$ labeled by the very same wave number~$k$ when the driving 
amplitude~$F_1$ goes to zero, such that, in particular, the band edges are 
labeled by $k=0$ and $k = \pm\pi/a$ for each~$F_1$. This is achieved by 
setting $k_0 = 0$, so that $k - A(t)/\hbar$ oscillates 
around~$k$~\cite{EckardtEtAl09,ArlinghausHolthaus11}. In the case of 
{\em unconventional\/} quasienergy bands with $r \neq 0$ the requirement of 
continuity does not suffice, since each representative of the quasienergy 
band~(\ref{eq:QTB}) collapses onto a rung of the Wannier-Stark ladder when 
$F_1$ vanishes while $F_0$ is held constant, since $J_r(0) = 0$ for $r \neq 0$. 
Then we stipulate $k_0 a = -r\varphi_0$, so that the band edges again obtain 
their usual labels when $J_r$ assumes nonzero 
values.

Finally, by inverting the transformation~(\ref{eq:DTR}) we obtain the 
spatio-temporal Bloch waves in the form of Eq.~(\ref{eq:STB}),
\begin{eqnarray}
	| \psi_k(t) \rangle & = & \sum_\ell 
	\exp\!\Big( \ri \big [k - A(t)/\hbar\big] \ell a \Big) | \ell \rangle  
\nonumber \\	& & \times	
	\exp\left( -\frac{\ri}{\hbar} \int_{t_0}^t \! \rd \tau \,
	E\big(k - A(\tau)/\hbar\big) \right) \; ; 
\label{eq:HOU}
\end{eqnarray}	          
observe that this $T$-periodic transformation does not affect the quasienergy
spectrum. In solid state physics, such states~(\ref{eq:HOU}) are known as 
``accelerated Bloch waves'', or Houston functions~\cite{Houston40}. If these 
functions already gave a complete description of the dynamics of ultracold 
atoms in tilted and periodically driven deep optical lattices, there would be 
no excitation at all: One merely would have to deal with wave 
packets made up from these Houston functions; such packets would then 
conform to the dispersion relation~(\ref{eq:QTB}) for all 
times~\cite{ArlinghausHolthaus11}. In reality, 
the forcing-induced coupling to ``higher'' bands  that have been neglected in 
this single-band approximation enables energy absorption, causing the atoms 
to finally escape from the lattice~\cite{WeinbergEtAl15,
StraeterEckardt16,ReitterEtAl17}, as will be explored in quantitative detail 
in the following section.

\section{Numerical calculations}
\label{sec:4}

In an approximate sense, wave packets of Houston functions~(\ref{eq:HOU}) 
may be regarded as initial states of the excitation process; the 
final states would then be provided by the Airy function-based analogs of the 
Gordon-Volkov wave functions constructed in Appendix~\ref{app:1}. Such an 
approach, while physically reasonable, still misses an important detail. A 
homogeneous force exerted on a periodic potential couples only states labeled 
by the same wave number~$k$; if there is only one single band, as assumed 
in the previous section, there is no coupling at all. In this regard, the Houston 
functions~(\ref{eq:HOU}) are trivial. In contrast, the lattice potential 
$V(x)$ possesses infinitely many energy bands, Bloch states with the 
same~$k$ being coupled by the driving force. Therefore, understanding the 
quantum mechanism that underlies the energy absorption by the atoms 
necessarily requires knowledge of the respective full quasienergy band 
structure.

Here we focus on a one-dimensional optical cosine lattice potential with 
depth~$V_0$,      
\begin{equation}
	V(x) = -\frac{V_0}{2} \cos(2 k_{\rm L} x) \; ,
\label{eq:OCL}	
\end{equation}
where $k_{\rm L}$ designates the wave number of the lattice-generating laser 
radiation, so that the lattice period is $a = \pi/k_{\rm L}$. The 
characteristic energy scale then is set by the single-particle recoil energy
\begin{equation}
	E_{\rm R} = \frac{\hbar^2 k_{\rm L}^2}{2M} \; ;
\end{equation}		
for instance, with $^{87}$Rb atoms in a lattice with $2\pi/k_{\rm L} = 
830$~nm, as employed in recent measurements of interband transitions by 
Weinberg {\em et al.\/}~\cite{WeinbergEtAl15}, one has 
$E_{\rm R}/(2\pi\hbar) = 3.329$~kHz.

For the sake of definiteness we also set $\varphi_0 = 0$ in the driving
force~(\ref{eq:ADF}); this is tantamount to a particular fixation of the moment
$t = 0$. In view of Eq.~(\ref{eq:POT}), this choice implies that both the 
velocity-gauge Schr\"odinger equation~(\ref{eq:TRE}) and the G\"oppert-Mayer
operators~(\ref{eq:GMO}) remain invariant under the temporal parity operation
\begin{equation}
	\mathcal{T} : \left\{ \begin{array}{l}
		t \, \to \, -t	\\
		k    \to    -k	\\
		\text{complex conjugation.}	
		\end{array} \right.
\end{equation}
Hence, one has $u_{n,-k}(x,t) = u_{n,k}^*(x,-t)$ and symmetric dispersion
relations, $\varepsilon_n(-k) = \varepsilon_n(k)$.		
	
For solving the eigenvalue problems~(\ref{eq:QEP}) numerically we resort 
to the strategy opened up by Eq.~(\ref{eq:STR}), implementing the periodic 
boundary conditions in space~(\ref{eq:PBS}) with the help of a basis set of 
trigonometric functions for computing $\widetilde{U}_k(T,0)$, while the 
periodic boundary conditions in time~(\ref{eq:TBC}) are accounted for by 
diagonalization of the one-cycle evolution operators $U_k(T,0)$ pertaining
to the length gauge. 

We now consider a fairly deep lattice, $V_0/E_{\rm R} = 30$, so that the 
tight-binding approximation should retain at least some meaning. The width 
of the lowest energy band then is $W/E_{\rm R} = 1.85 \times 10^{-3}$, while 
the energy gap $\Delta$ between this ground-state Bloch band and the first 
excited band figures as $\Delta/E_{\rm R} = 9.80$. We apply a static force of 
strength $F_0 a/E_{\rm R} = 0.072$, so that the well-to-well tilt is much smaller 
than the lattice depth. Moreover, we apply resonant time-periodic forcing 
with $r = 1$, as depicted symbolically in Fig.~\ref{F_1}, so that the spacing 
between the rungs of the Wannier-Stark ladder equals the energy quantum of the 
drive, $F_0a = \hbar\omega$. This also implies $\Delta/(\hbar\omega) = 136.1$,
so that the energy of about 136~``photons'' is required to bridge the 
gap~$\Delta$, meaning that interband excitation by multiphoton-like 
transitions~\cite{WeinbergEtAl15,StraeterEckardt16,ReitterEtAl17} is strongly
suppressed.

\begin{figure}[t]
\begin{center}
\includegraphics[width = 0.9\linewidth]{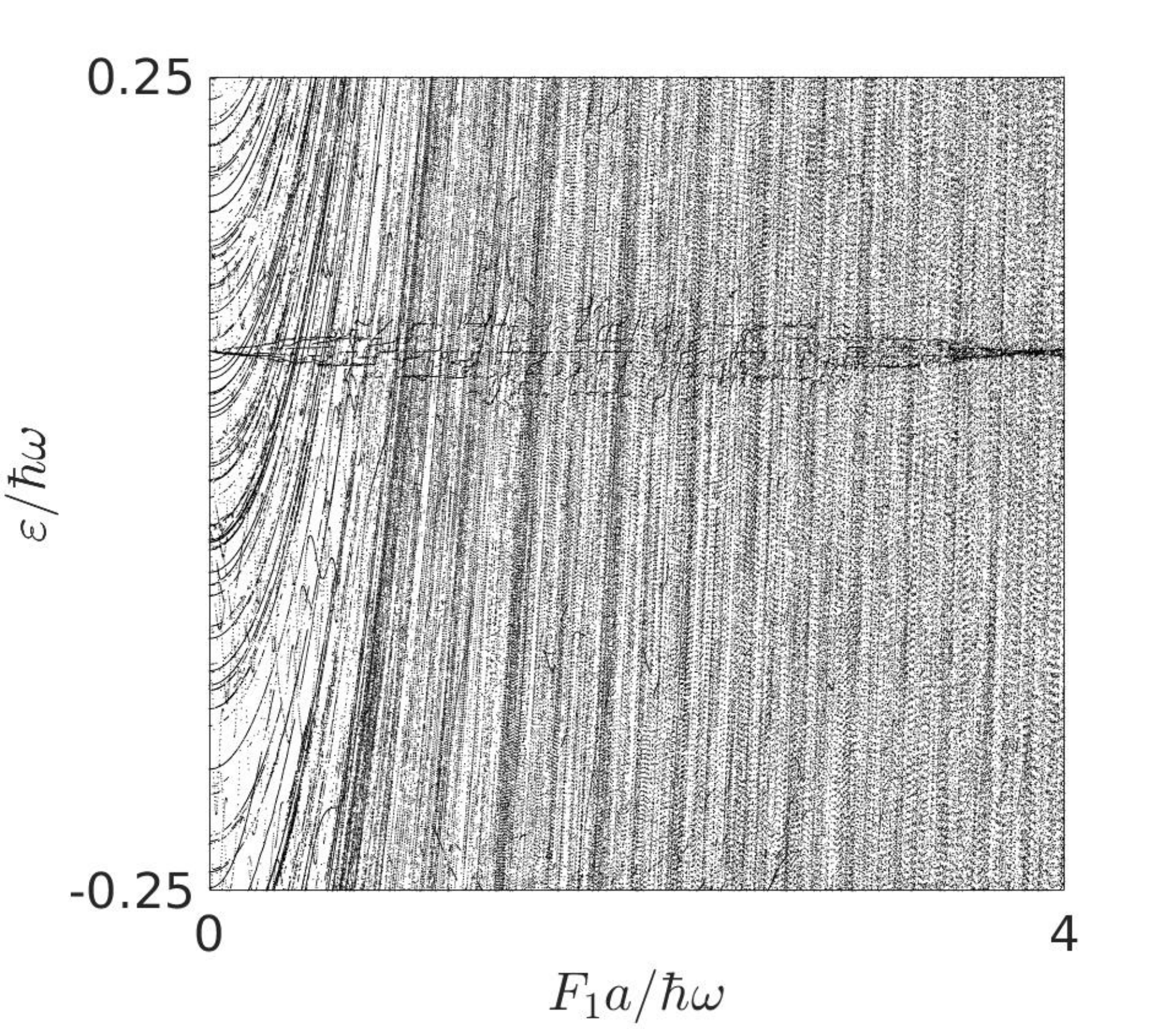}
\end{center}
\caption{Half a Brillouin zone of quasienergies for an optical 
	lattice with depth $V_0/E_{\rm R} = 30$ subjected to both a static 
	bias with strength $F_0 a/E_{\rm R} = 0.072$ and resonant periodic 
	driving with $r = 1$, vs.\ the scaled amplitude $F_1 a/(\hbar\omega)$ 
	of the drive. Quasienergies belonging to $k/k_{\rm L} = 0.0$, $0.1$,
	\ldots, $1.0$ have been superimposed, in order to bring out the remnant
	of the tight-binding band~(\ref{eq:QTB}). Observe that this band should
	become flat when $F_1 a/(\hbar\omega)$ approaches the zero of the 
	Bessel function $J_1(z)$ at $z \approx 3.83$. The heavy background is 
	due to Floquet states that are hardly affected by the lattice potential, 
	so that their quasienergies are given by Eq.~(\ref{eq:PES}).} 	        
\label{F_2}
\end{figure}

In Fig.~\ref{F_2} we display half a Brillouin zone of quasienergies for this
system, $-1/4 \leq \varepsilon_n(k)/(\hbar\omega) < +1/4$, vs.\ the scaled 
amplitude $F_1 a/(\hbar\omega)$ of the drive. Here we have superimposed the 
results obtained for $11$ equally spaced wave numbers comprising half a 
reciprocal-lattice Brillouin zone, $k/k_{\rm L} = 0.0$, $0.1$, \ldots, $1.0$.
Although this unorthodox way of combining quasienergy eigenvalues 
makes this figure rather cluttered, it renders the remnant of the tight-banding 
band~(\ref{eq:QTB}) originating from the ground-state energy band of the 
undriven lattice clearly visible against a ``background'' of quasienergies that 
increase quadratically with the driving amplitude. This is precisely what is 
expected for Floquet states that do not ``feel'' the lattice potential, as 
constructed in Appendix~\ref{app:1}, so that the variation of their 
quasienergies with the driving amplitude is determined by the ponderomotive 
energy in accordance with Eq.~(\ref{eq:PES}). Because of the Brillouin zone 
structure of the quasienergy spectrum, which requires that there is one 
representative of the quasienergy of each and every state in each and every 
Brillouin zone, this background is virtually ``dense''. It is of key importance 
to realize how this dense background affects the tightly bound Houston 
states: According to the von Neumann-Wigner theorem~\cite{NeumannWigner29}, 
eigenvalues of Hermitian operators belonging to the same symmetry class should 
not cross, but exhibit avoided crossings, to the effect that the anticrossing 
eigenstates hybridize. Applied to the ground-state quasienergy band~(\ref{eq:QTB}) 
that is hit virtually everywhere by the quasienergies~(\ref{eq:PES}) stemming from 
high-energetic above-barrier states, this means that the original band is disrupted 
by a vast plethora of narrow avoided crossings, the width $\delta \varepsilon$ 
of which determines the time scale $\hbar/\delta\varepsilon$ for populating the 
respective anticrossing state, {\em i.e.\/}, for energy absorption.

\begin{figure}[t]
\begin{center}
\includegraphics[width = 0.9\linewidth]{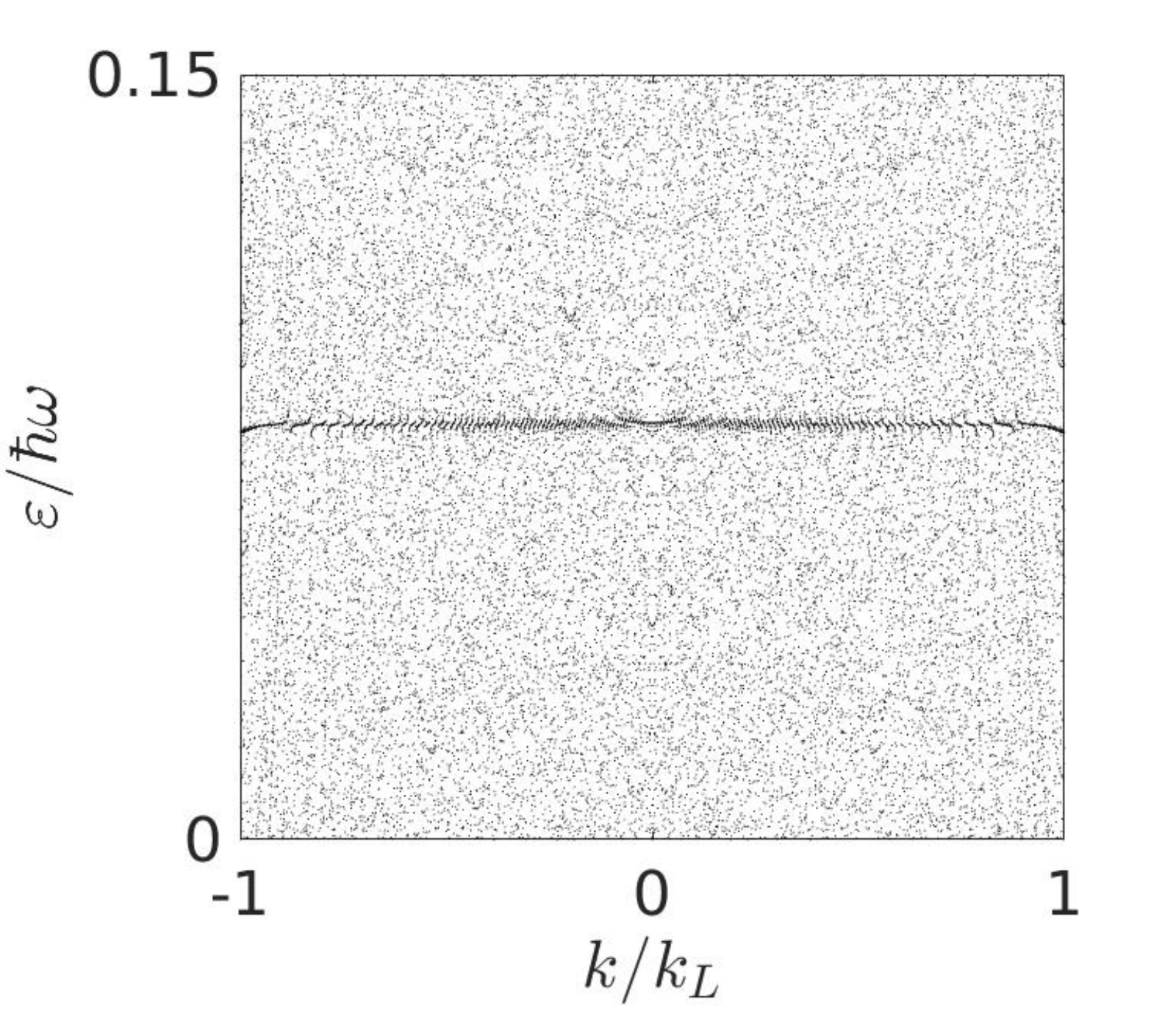}
\end{center}
\caption{Dispersion relation of the unconventional, non-Kohnian quasienergy 
	band considered in Fig.~\ref{F_2} for vanishing driving amplitude, 
	$F_1 a/(\hbar\omega) = 0$. The perfect horizontal line expected for 
	an isolated tight-binding band~(\ref{eq:QTB}) is perforated by a 
	multitude of avoided crossings with the background states that here 
	show up as dots scattered in a seemingly ir\-regular but homogeneous 
	manner.} 
\label{F_3}
\end{figure}

\begin{figure}[t]
\begin{center}
\includegraphics[width = 0.9\linewidth]{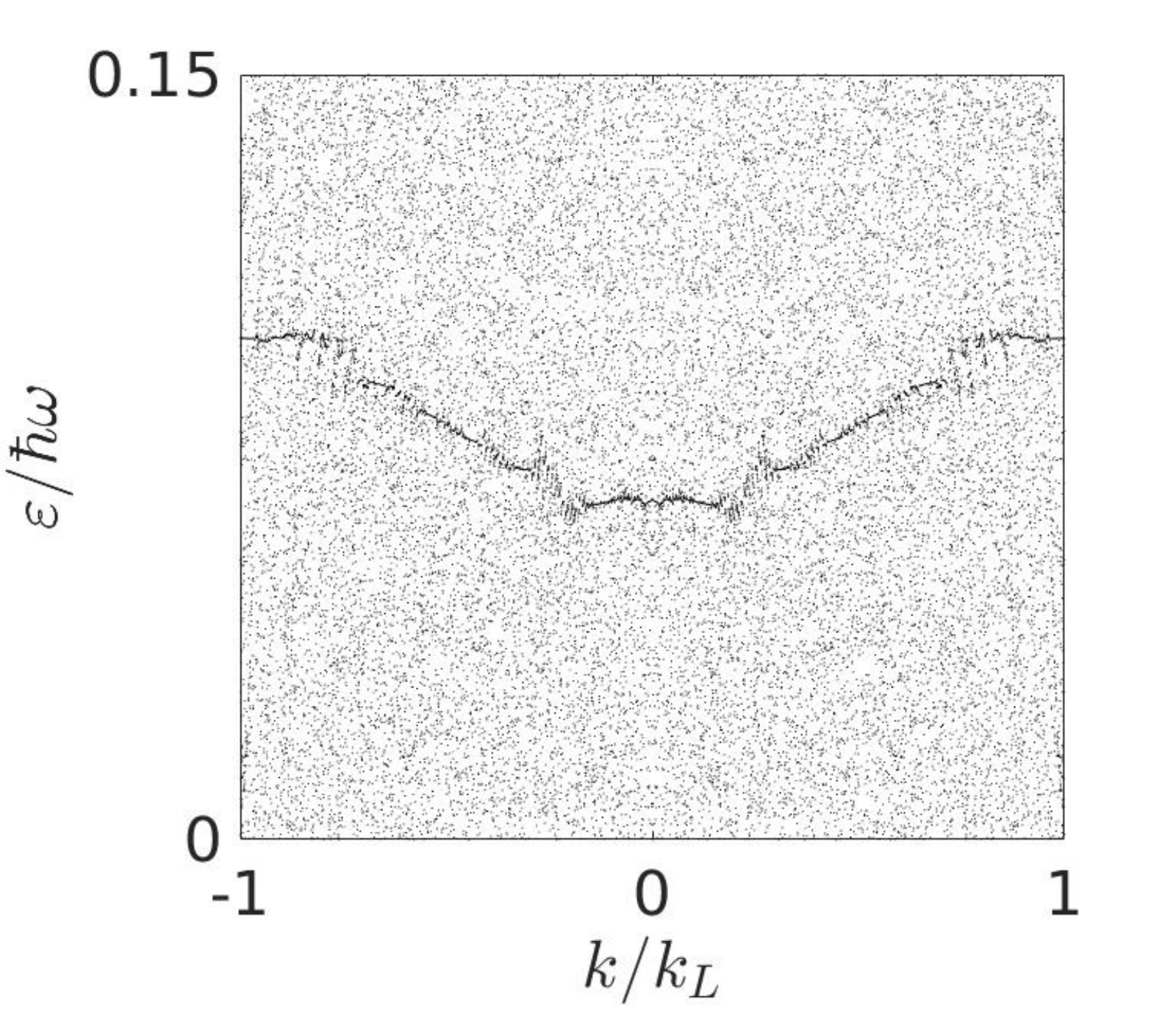}
\end{center}
\caption{As Fig.~\ref{F_3}, for $F_1 a/(\hbar\omega) = 1.0$.
	With $J_1(1.0) \approx 0.440$, the isolated band~(\ref{eq:QTB}) 
	acquires a cosine shape that is strongly distorted here, being 
	masked by dense tangles of avoided quasienergy crossings.} 
\label{F_4}
\end{figure}

\begin{figure}[ht]
\begin{center}
\includegraphics[width = 0.9\linewidth]{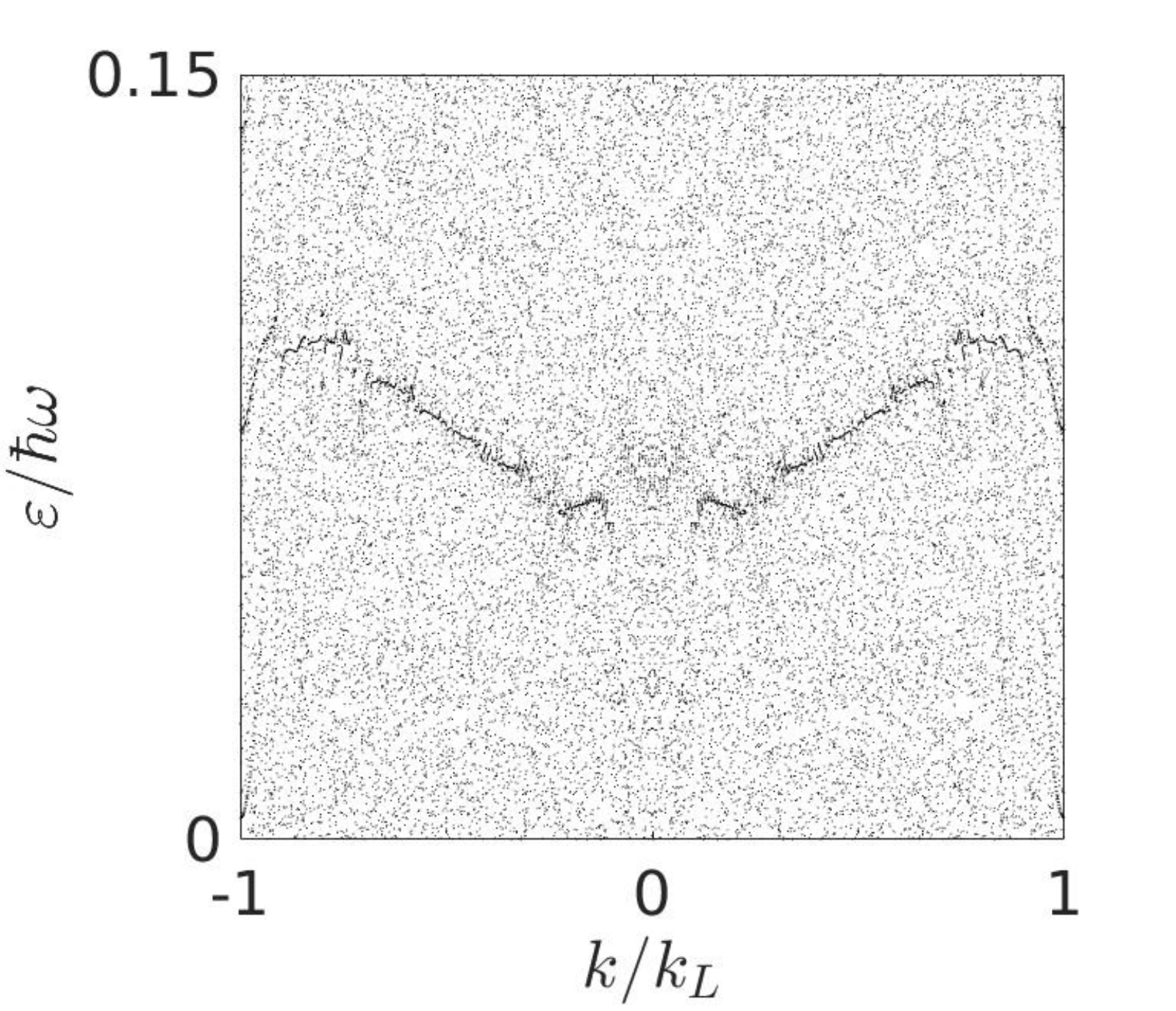}
\end{center}
\caption{As Fig.~\ref{F_3}, for $F_1 a/(\hbar\omega) = 2.5$.
	Note that $J_1(2.5) \approx 0.497$, so that the width of this 
	non-Kohnian band is comparable to that in Fig.~\ref{F_4}.} 
\label{F_5}
\end{figure}

\begin{figure}[t]
\begin{center}
\includegraphics[width = 0.9\linewidth]{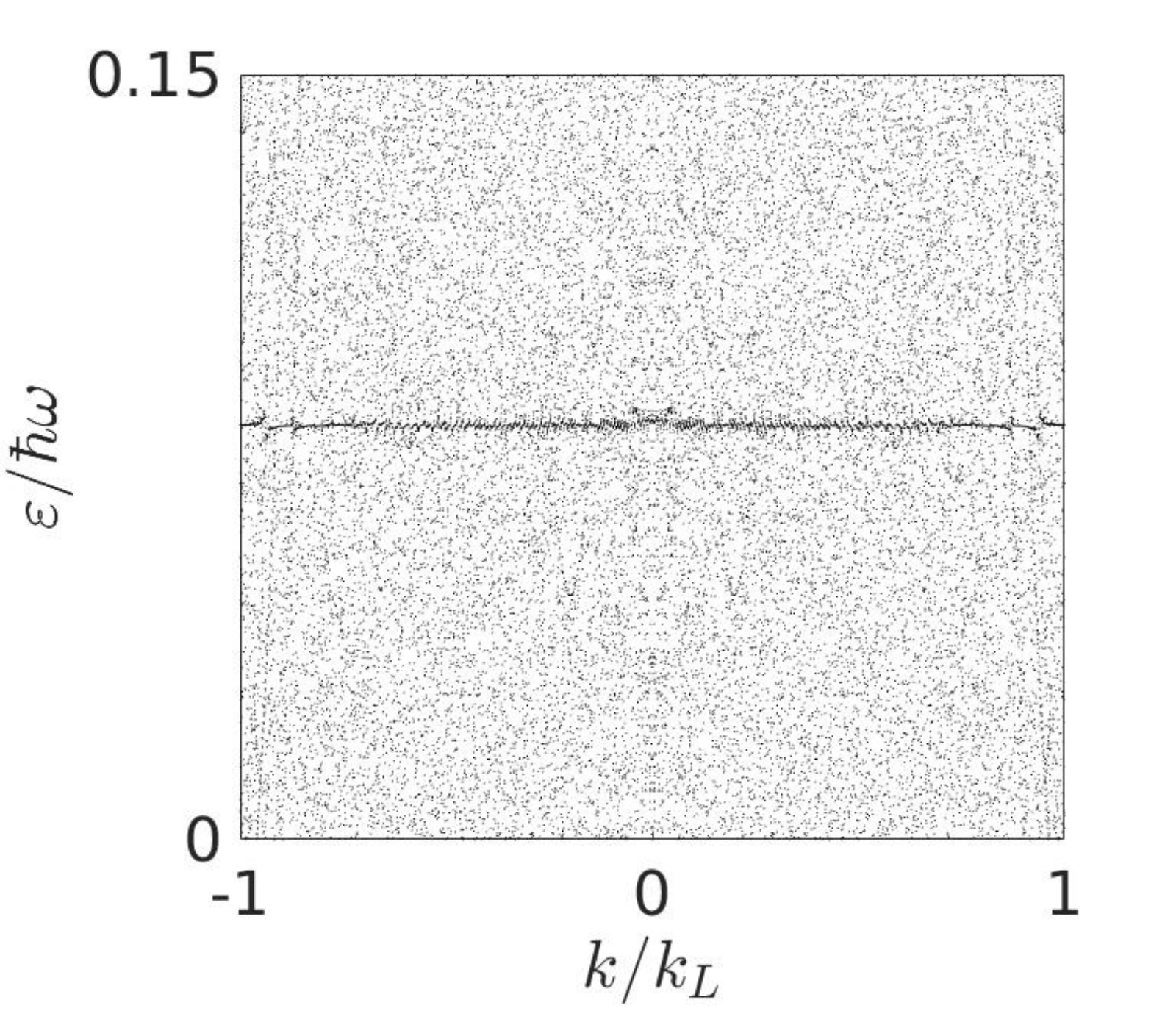}
\end{center}
\caption{As Fig.~\ref{F_3}, for $F_1 a/(\hbar\omega) = 3.83$, close to the 
	first nontrivial zero of $J_1(z)$. Instead of being perfectly flat, 
	the ground-state quasienergy band is lacerated by numerous 
	anti\-crossings.} 
\label{F_6}
\end{figure}

This scenario is visualized in an even more drastic manner if one plots 
the quasienergy dispersion relations $\varepsilon_n(k)$ at fixed driving 
amplitude, as would be appropriate for the description of an experiment 
aiming at the system's response to long-time ac-dc forcing. In Figs.~\ref{F_3}, 
\ref{F_4}, \ref{F_5}, and \ref{F_6} we depict these relations for 
$F_1 a/(\hbar\omega) = 0.0$, $1.0$, $2.5$, and $3.83$, respectively. 
With these figures we do not intend to resolve all details of the 
actual quasienergy spectra, which would be a hopeless enterprise. Since 
we are dealing with point spectra that are ``dense'' even in the rigorous 
mathematical sense, any apparent avoided crossing will be broken itself by 
infinitely many finer ones that, again, are subject to perforation, and so on, 
resulting in uncomputable tangles of avoided crossings on all scales. On the 
other hand, numerical computations with finite basis sets procure eigenvalues 
that provide a stable density upon increase of the basis size, concentrating 
at the quasienergies of the physical states. That is, here we aim at a visualisation 
of the spectral concentration of the eigenvalues, a concept that was invented 
by Titchmarsh~\cite{Titchmarsh51} and applied in particular to Floquet 
operators in Ref.~\cite{Howland90}. 

The first of these examples, depicted in Fig.~\ref{F_3}, refers to a tightly 
bound Bloch particle under constant dc bias only~\cite{Zak96}: With $J_1(0) = 0$, 
each representative of the ideal tight-binding band~(\ref{eq:QTB}) aquires 
zero width at $F_1 a/(\hbar\omega) = 0$, collapsing onto a rung of the 
Wannier-Stark ladder, so that one would obtain a horizontal quasienergy line 
for such an isolated band. In contrast, unavoidable coupling to the background 
states effectuates a strongly lacerated line in Fig.~\ref{F_3} that, on a coarse 
scale, appears as if it had a finite width, as determined by the
density of eigenvalues.

The isolated quasienergy band~(\ref{eq:QTB}) adopts a cosine shape when
the scaled driving amplitude $F_1 a/(\hbar\omega) = z$ is shifted to moderate
nonzero values. However, this shape may barely be attainable in real tilted 
and resonantly driven optical lattices, as shown in Fig.~\ref{F_4} for 
$z = 1.0$, invoking $J_1(1.0) \approx 0.440$, and in Fig.~\ref{F_5} for 
$z = 2.5$, with $J_1(2.5) \approx 0.497$.

A noteworthy case is shown in Fig.~\ref{F_6}: The Bessel function $J_1(z)$
possesses a zero at $z \approx 3.83$, so that the ideal band~(\ref{eq:QTB}) 
becomes perfectly flat at this value of the scaled driving amplitude. But 
again, coupling to the background states gives rise to lots and lots of tiny 
avoided quasienergy crossings, opening up the excitation channels.   

The task now is to deduce characteristic escape times for ultracold 
atoms initially stored in the ground-state band from these quasienergy spectra. 
To this end, we adapt a particular approach for calculating decay widths of 
metastable states, known as box stabilization method~\cite{MandelshtamEtAl93,
MakarovEtAl94}. As sketched briefly in Appendix~\ref{app:2}, one may 
``stabilize'' a decaying system by confining it to a box of size~$L$, and 
compute the square-integrable eigenstates of the confined system. Not only does
the precursor of a decaying state manifest itself clearly in a stabilization 
diagram, that is, in a plot of the energy eigenvalues vs.\ the box size~$L$, 
but appropriate binning of these eigenvalues even allows one to extract both 
the resonance energy and the lifetime of such decaying states. There is an 
evident conceptual similarity between such stabilization diagrams, as 
exemplified by Fig.~\ref{F_B1} in Appendix~\ref{app:2}, and the previous 
Figs.~\ref{F_3} -- \ref{F_6}, both types of plots visualizing the coupling of 
tightly bound states to a background effectuating their decay. In 
both cases, the typical width of the avoided crossings signaling this coupling 
encodes the lifetime of an initial wave packet. Thus, we suggest to construct 
empirical quasienergy densities of Floquet-Bloch states for ultracold atoms in 
periodically driven optical lattices with the help of histograms computed by 
counting the number of quasienergy eigenvalues that fall into bins of suitable
size $\Delta\varepsilon$ while $k/k_{\rm L}$ ranges over one Brillouin zone of
the reciprocal lattice. Perforated flat quasienergy bands then give rise to
a Lorentzian peak with a width $\Gamma$ that measures the average width
of the anticrossings, so that $\tau = \hbar/\Gamma$ is an estimate of the 
desired escape time. There is an obvious caveat, though: 
In the case of the original box stabilization method the box size~$L$ is an 
unphysical parameter that may be varied in an almost arbitrary manner; 
in our case the parameter $k/k_{\rm L}$ has a physical meaning, and is 
restricted to the interval $-1 \le k/k_{\rm L} < +1$.

\begin{figure}[t]
\begin{center}
\includegraphics[width = 0.9\linewidth]{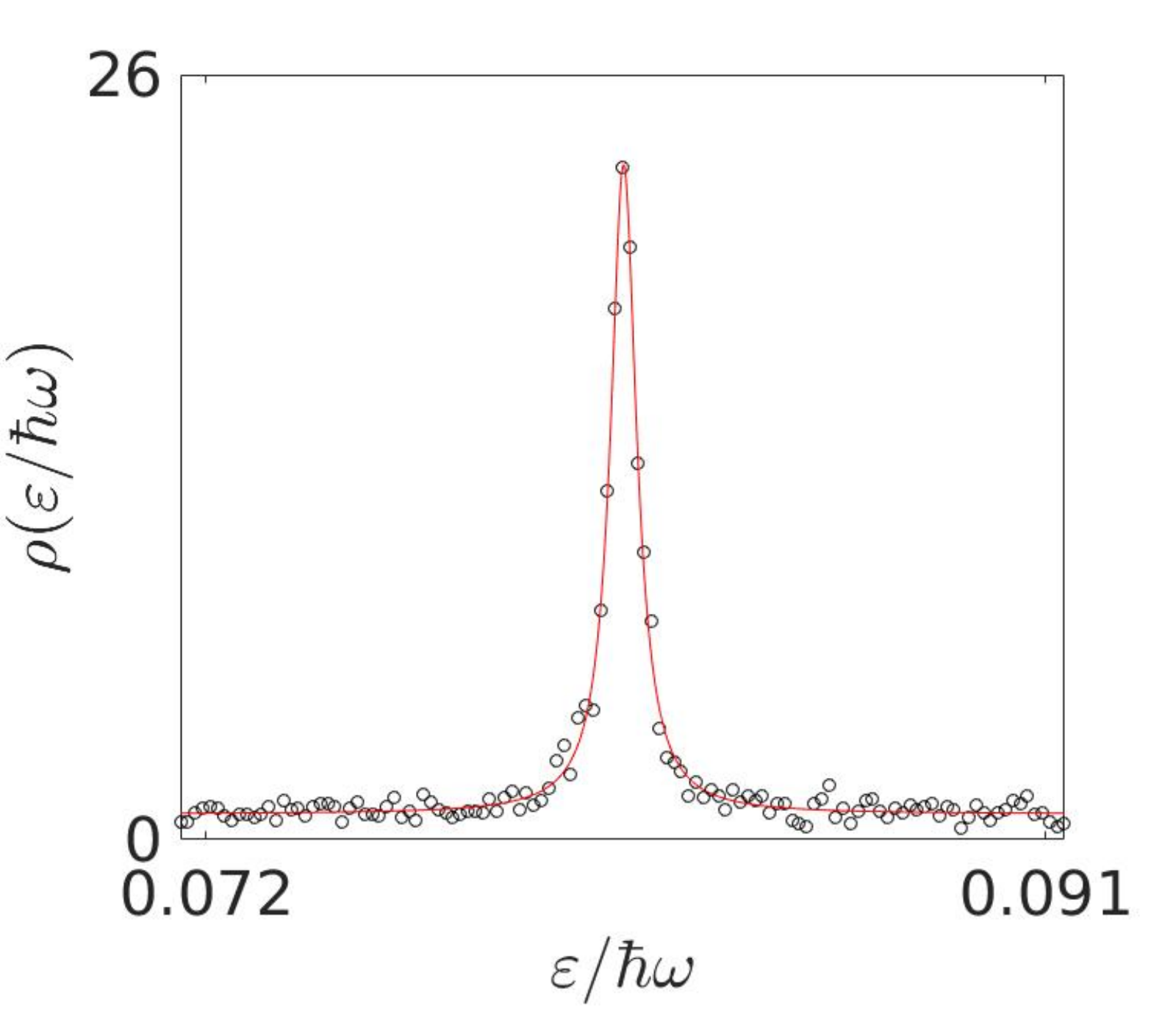}
\end{center}
\caption{(Color online) Empirical density of Floquet-Bloch states obtained 
	by binning the quasienergy spectrum shown in Fig.~\ref{F_3}, where
	$F_1 a/(\hbar\omega) = 0$. Circles indicate histogram data; the full 
	line results from their fit to a Lorentzian. The width of the peak is 
	$\Gamma/(\hbar\omega) \approx 7.0 \times 10^{-4}$ (FWHM), 
	translating into the lifetime $\tau/T \approx 226$ of wave packets 
	in the tilted, but undriven lattice.}  
\label{F_7}
\end{figure}

Figure~\ref{F_7} depicts the empirical density of states~$\rho$ obtained in
this manner from the spectrum shown in Fig.~\ref{F_3}, that is, for vanishing 
driving amplitude. Evidently the fit to a Lorentzian is quite good, providing
the width $\Gamma/(\hbar\omega) \approx 7.0 \times 10^{-4}$ (FWHM), and thus 
yielding the lifetime $\tau/T \approx 226$ of a wave packet in the tilted, 
but yet undriven lattice. For the sake of comparison with the following cases, 
in which the periodic drive is present, the reference time is the cycle length 
$T = 2\pi/\omega$.  
  
\begin{figure}[t]
\begin{center}
\includegraphics[width = 0.9\linewidth]{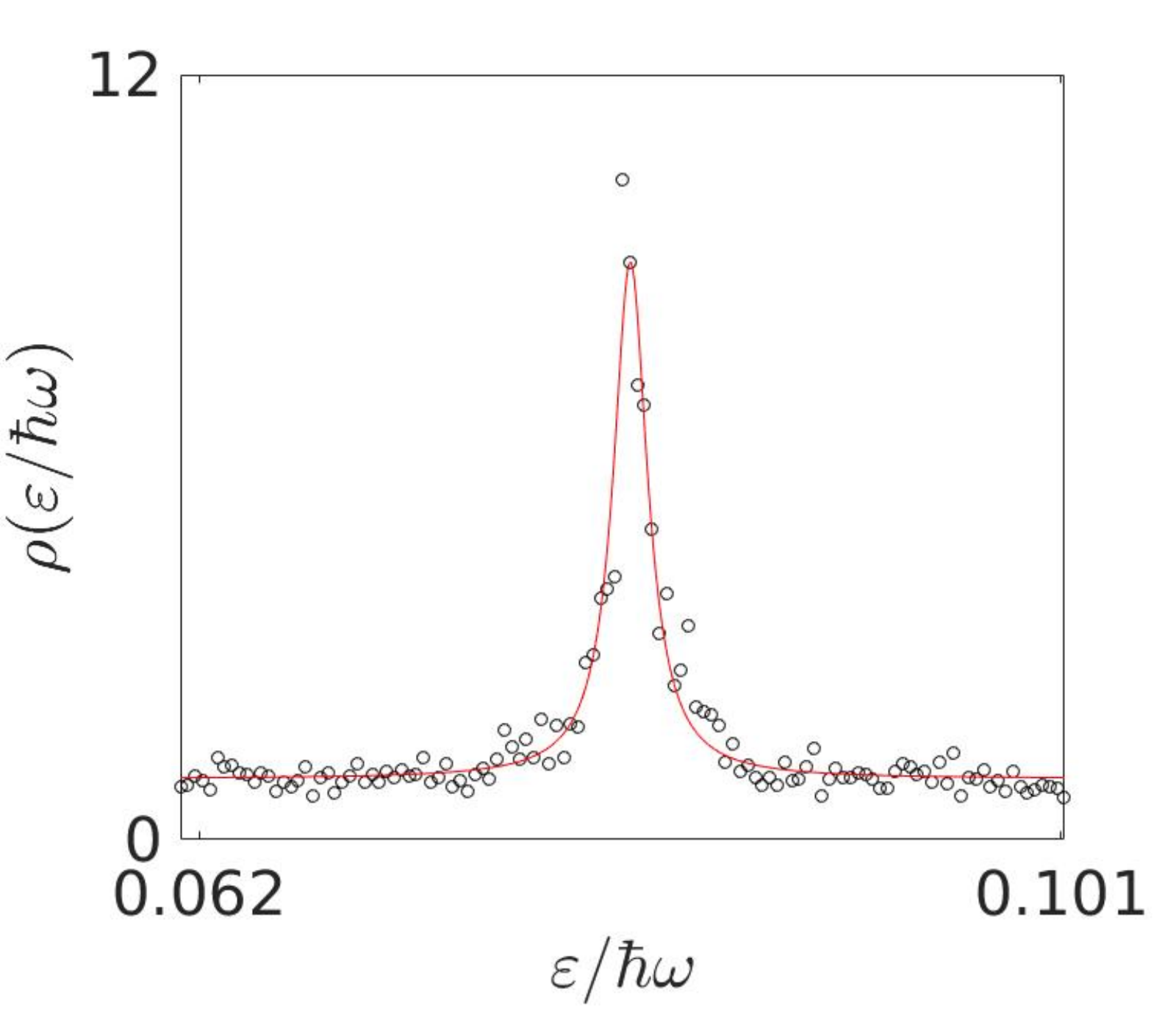}
\end{center}
\caption{(Color online) Empirical density of Floquet-Bloch states obtained 
	by binning the quasienergy spectrum shown in Fig.~\ref{F_4}, 
	where $F_1a/(\hbar\omega) = 1.0$, after correcting for the ideal 
	cosine band~(\ref{eq:QTB}) with suitable effective width~$W^*$. The 
	width of the fitted Lorentzian is $\Gamma/(\hbar\omega) \approx 
	1.9 \times 10^{-3}$, translating into the escape time 
	$\tau/T \approx 84$.}  
\label{F_8}
\end{figure}

\begin{figure}[ht]
\begin{center}
\includegraphics[width = 0.9\linewidth]{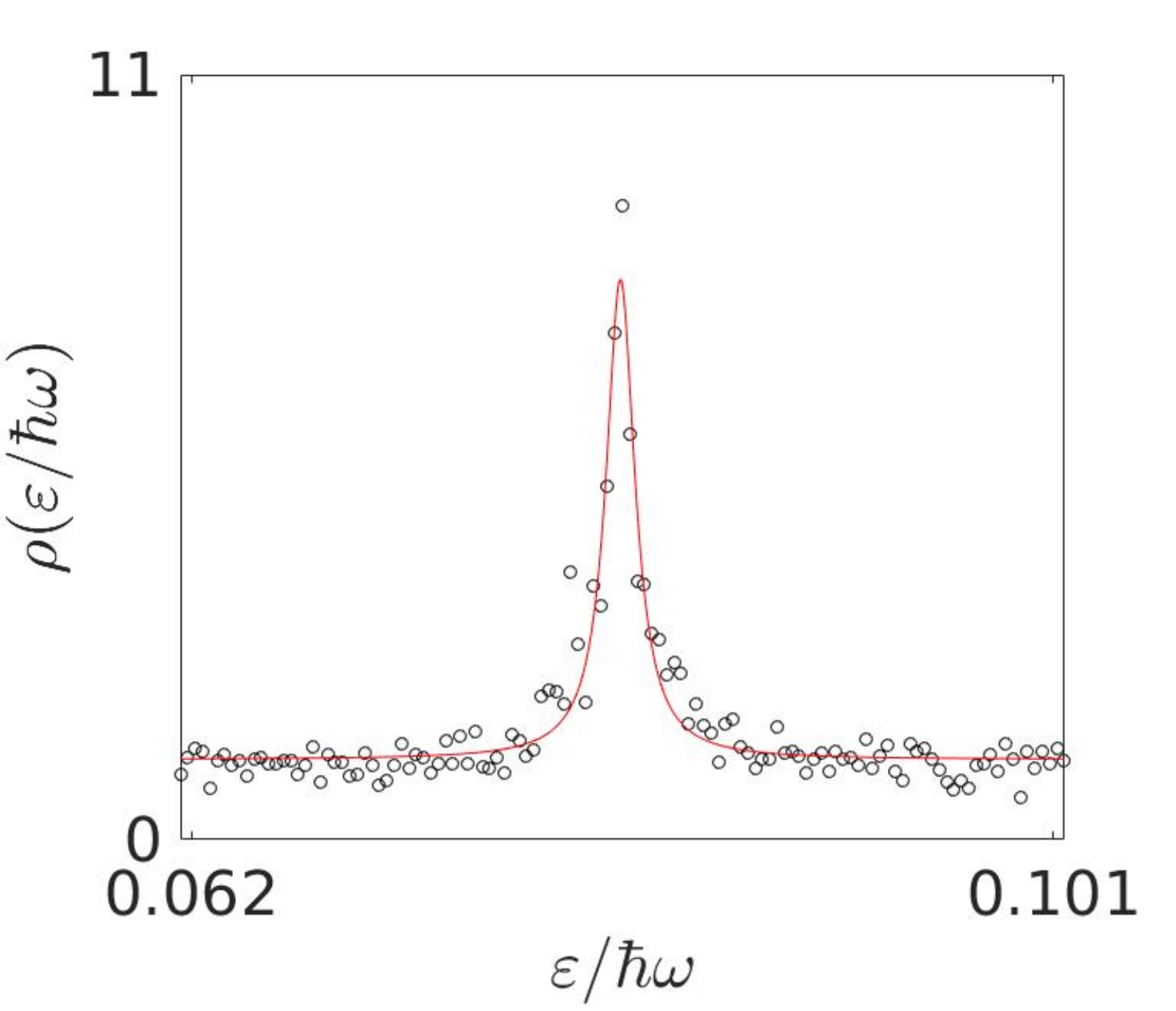}
\end{center}
\caption{(Color online) Empirical density of Floquet-Bloch states obtained 
	by binning the quasienergy spectrum shown in Fig.~\ref{F_5}, 
	where $F_1a/(\hbar\omega) = 2.5$, after correcting for the ideal 
	cosine band~(\ref{eq:QTB}) with suitable effective width~$W^*$. The 
	width of the fitted Lorentzian is $\Gamma/(\hbar\omega) \approx 
	1.6 \times 10^{-3}$, translating into the escape time 
	$\tau/T \approx 102$.}  
\label{F_9}
\end{figure}

Before applying the same technique to the situations previously considered 
in Figs.~\ref{F_4} and~\ref{F_5}, we need to take out the variation of the 
perforated quasienergy bands with the wavenumber $k/k_{\rm L}$ that is already
exhibited by the ideal band~(\ref{eq:QTB}). Remarkably, it turns out that the 
smoothed quasienergy bands are well described by a cosine band~(\ref{eq:QTB}) 
with a ``renormalized'' width parameter $W^*$ substantially larger than the 
original band width~$W$. Therefore, we subtract these renormalized dispersion 
relations from the numerically computed eigenvalues~$\varepsilon_n(k)$ before 
doing the histogram count. We wish to stress that at present this 
procedure is based on a working hypothesis that may well require further 
thoughts, and might inspire future researchers to devise improvements.  
The results are displayed in Fig.~\ref{F_8} for 
$F_1a/(\hbar\omega) = 1.0$ and $W^*/W = 2.83$, giving the escape
time $\tau/T \approx 84$, and in Fig.~\ref{F_9} for much stronger driving, 
$F_1a/(\hbar\omega) = 2.5$ and $W^*/W = 2.83$, yielding $\tau/T \approx 102$.   
In both cases the fit to a Lorentzian is not as good as in Fig.~\ref{F_7},
possibly because the {\em ad hoc\/} correction procedure is not yet optimal. 
Nonetheless, the expected reduction of $\tau$ due to the action of the 
periodic drive can be deduced clearly.

\begin{figure}[t]
\begin{center}
\includegraphics[width = 0.9\linewidth]{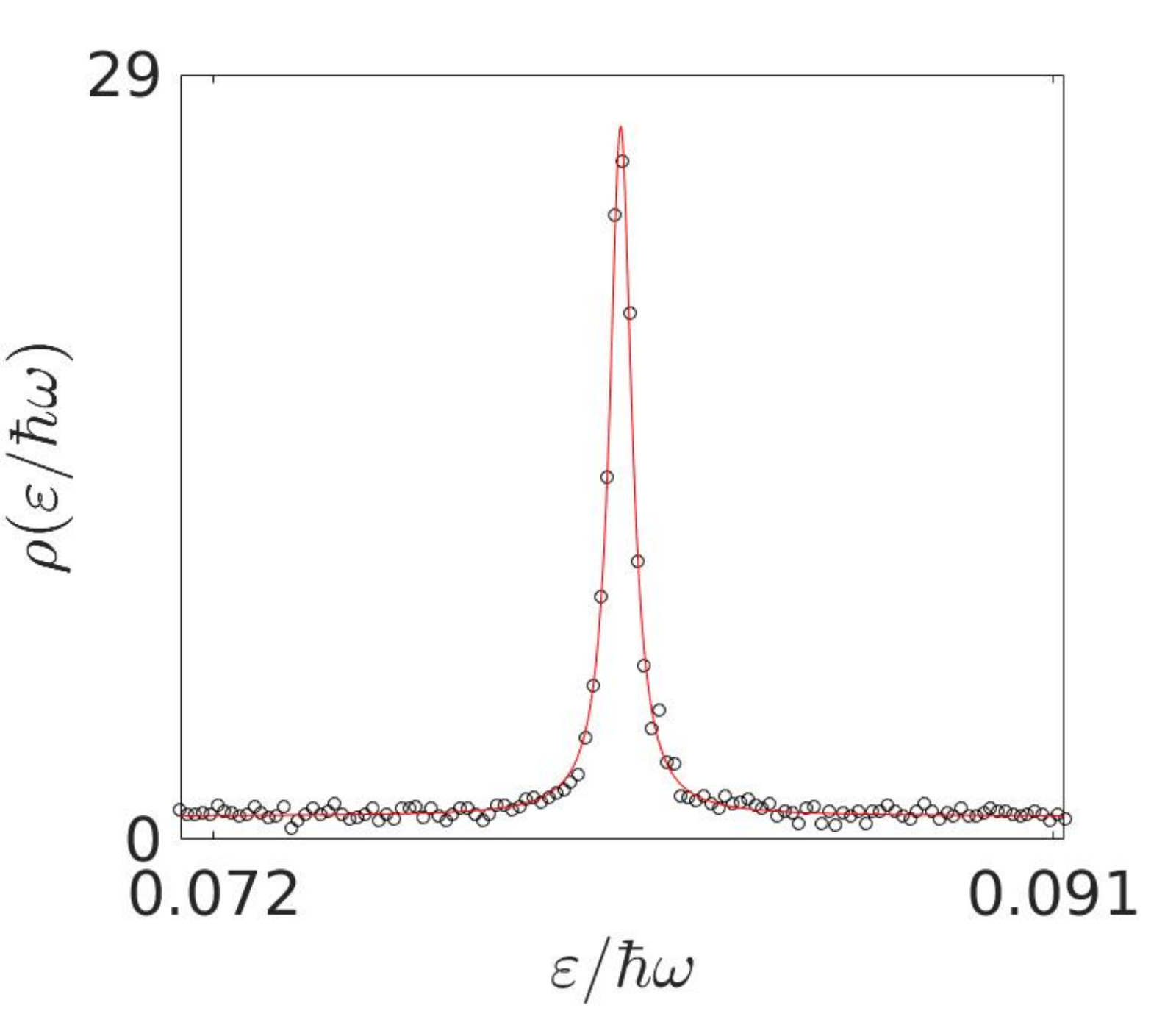}
\end{center}
\caption{(Color online) Empirical density of Floquet-Bloch states obtained 
	by binning the quasienergy spectrum shown in Fig.~\ref{F_6}, where 
	$F_1a/(\hbar\omega) = 3.83$, so that the ideal band~(\ref{eq:QTB})
	is flat. The width of the fitted Lorentzian is $\Gamma/(\hbar\omega) 
	\approx 5.9 \times 10^{-4}$, translating into the escape 
	time $\tau/T \approx 267$ of wave packets in the tilted and strongly 
	driven optical lattice.}  
\label{F_10}
\end{figure}

The special values of the scaled driving amplitude $F_1a/(\hbar\omega)$ 
that correspond to a zero of the Bessel function $J_r(z)$ bear particular
significance, since they make the ideal tight-binding band~(\ref{eq:QTB}) 
``collapse''~\cite{Holthaus92}. Hence, they allow for binning entirely without 
band shape correction, relieving us of the awkwardness of having to invoke 
plausible but unproven hypotheses, such as implied by an effective band width 
$W^*$. In Fig.~\ref{F_10} we display the density of Floquet-Bloch states for 
$r = 1$ and  $F_1 a/(\hbar\omega) = 3.83$, as obtained in this unbiased manner 
from the spectrum shown in Fig.~\ref{F_6}. Indeed, here the fit to a Lorentzian 
is fairly convincing again. And there is a quite surprising observation to be 
made: The escape time $\tau/T \approx 267$ deduced here for the 
periodically driven lattice is even longer than what has been found in 
Fig.~\ref{F_7} in the absence of the ac force, although the driving amplitude 
is rather large --- with $\hbar\omega/E_{\rm R} = 0.072$ we have 
$F_1 a/E_{\rm R} = 0.276$, almost four times as large as the static 
well-to-well tilt $F_0 a/E_{\rm R}$~!

\section{Discussion}
\label{sec:5}

The classic description of Bloch waves and energy bands pertaining to
one-dimensional periodic potentials as branches of multivalued analytic and 
periodic functions of a complex wave number~$k$ given by Kohn~\cite{Kohn59} 
is based on the proposition that the energy bands do not intersect. When 
dealing with quasienergy bands, this proposition is violated to the extreme: 
When reduced to the quasienergy Brillouin zone, quasienergy bands originating 
from the infinitely many, well separated energy bands of an undriven lattice 
necessarily are projected onto each other. This results in tangles of 
anticrossings that, as a matter of principle, are even uncomputably dense.
This well-understood feature reflects the fact that the quasienergy eigenvalue 
problem~(\ref{eq:QEP}) with the boundary conditions~(\ref{eq:SBC}) 
and~(\ref{eq:TBC}) in space and time is {\em much\/} more difficult 
mathematically than the usual textbook energy eigenvalue problem for Bloch 
waves with periodic boundary conditions in space only~\cite{AshcroftMermin76}. 
In this sense, Figs.~\ref{F_3} -- \ref{F_6} provide striking, yet natural 
examples of non-Kohnian bands. From an experimental viewpoint, ultracold 
atoms in periodically driven optical lattices may provide more convenient and 
more controllable access to the physics of such non-Kohnian bands 
than conventional solid-state materials exposed to laser 
radiation~\cite{KruchininEtAl18}.
 
From a conceptual viewpoint, the escape scenario we have studied 
in this paper --- ultracold atoms in an optical lattice that is both tilted 
by a static dc component of an externally applied force, and time-periodically 
driven by an ac component ---, may be thought of as a particular variant of 
photon-assisted tunneling. Indeed, atoms in a tilted lattice would eventually 
tunnel out of the lattice into a continuum of final states 
even without the presence of an additional time-periodic 
force~\cite{Avron82,GlueckEtAl02}. Yet, it needs to be realized that the 
static dc component of the force is not really necessary for conjuring the 
energy absorption scheme discussed here: The very same elements 
--- the appearance of non-Kohnian quasienergy bands, their perforation due to 
narrow avoided crossings with high-energetic above-barrier states, the 
connection between the escape times and the typical width of these 
avoided quasienergy crossings --- also show up when the force does not contain 
a dc contribution, so that tunneling is disabled. In that case there are stable 
energy bands when the ac force is switched off, and the escape 
times may be substantially longer~\cite{Holthaus16}. 

It might also be worthwhile to point out that the present excitation 
mechanism is closely related to a mechanism that had been advocated some time 
ago for explaining certain features that had been observed experimentally in the 
microwave ionization of highly excited hydrogen atoms. In such 
experiments, a beam of hydrogen atoms in a Rydberg state is sent through a 
microwave resonator~\cite{BayfieldKoch74}. Since the microwave  field amplitude 
therein is comparable to the binding Coulomb field experienced by a highly excited 
electron, the atoms gain energy within the resonator, whereupon the ionization yield 
is recorded as a function of the field strength.  In some cases, for instance for initial 
Rydberg states with principal quantum numbers $n_0 = 37$ or $n_0 = 38$ under 
microwave irradiation with frequency~$\omega/(2\pi) = 9.92$~GHz, this ionization 
curve is non-monotonous, such that an initial increase of the ionization yield 
upon increase of the driving amplitude is followed by a decrease of the yield 
when the field strength is enhanced further, so that the ionization curve 
exhibits a well-pronounced bump~\cite{RichardsEtAl89}. This finding posseses 
an interpretation akin to the escape scenario that is elaborated on in the present work: 
When the atoms enter the resonator, they encounter a fringe field with an amplitude 
that increases slowly on the time scale of a single microwave cycle $T = 2\pi/\omega$. 
Now the electronic quasienergy eigenvalues, considered as functions of, {\em e.g.\/}, 
the amplitude and the frequency of the driving field, form quasienergy surfaces that 
enable adiabatic following  in a manner analogous to adiabatic motion on 
Born-Oppenheimer surfaces in molecular physics.  Hence, when the atoms pass the 
resonator fringe field, they are adiabatically transmuted into the Floquet states that are 
connected to the initial Rydberg states, to the effect that the Rydberg Floquet states 
that are actually experimented with within the  resonator are prepared in its fringe 
field~\cite{BreuerHolthaus89}. Due to the multitude of Rydberg states, these 
quasienergy surfaces are heavily punctured by avoided crossings, similar to the 
unconventional quasienergy bands. With most of these avoided crossings being 
negligibly small, they are traversed by Landau-Zener-type transitions when 
reacting to the slowly increasing fringe field, so that highly diabatic motion on 
strongly perforated quasienergy surfaces can be viewed as effectively adiabatic 
motion on coarse-grained, smooth surfaces. If then this pseudo-adiabatic transport 
ends up in a regime of field strengths where the avoided crossings are somewhat wider 
than they are at even stronger fields, more energy is absorbed at lower amplitudes 
than at highers, effectuating a non-monotonous ionzation curve~\cite{BreuerEtAl89}. 
Hence, these early experimental observations~\cite{RichardsEtAl89} may constitute 
another signature of the energy absorption scheme that is studied in the present 
paper in a rather different context.

For demonstration purposes we have chosen an optical lattice with a depth 
of thirty recoil energies in our present theoretical study, 
$V_0/E_{\rm R} = 30$, in order to make easy connection with the Houston 
states~(\ref{eq:HOU}) emerging in the tight-binding limit. This choice 
implies that the driving frequency $\omega/(2\pi)$ would be rather low; 
with $E_{\rm R}/(2\pi\hbar) = 3.329$~kHz, for example~\cite{WeinbergEtAl15}, 
our parameter $\hbar\omega/E_{\rm R} = 0.072$ amounts to 
$\omega/(2\pi) = 239.7$~Hz. However, it would be feasible to employ more 
shallow lattices in laboratory practice. 

A most remarkable finding stems from a comparison of our Figs.~\ref{F_7} 
and~\ref{F_10}: In the presence of a truly strong time-periodic force, which 
naively might be expected to always act in a de-stabilizing manner, ultracold 
atoms in a tilted optical lattice can actually be longer-lived, {\em i.e.\/}, 
more stable than they would be in the absence of that force. In other words, 
besides the expected photon-assisted tunneling there also exists a novel kind 
of photon-deactivated tunneling mechanism. This is a ``hard'' prediction 
that now is open to experimental verification, and may have interesting 
parallels to the intensely debated stabilization of atoms in strong laser 
fields~\cite{Gavrila02,PopovEtAl03,ZimmermannEtAl18}. Other unexpected 
phenomena might come into play when interparticle interaction has to be 
accounted for. Thus, we dare to conclude that the study of non-Kohnian 
quasienergy bands is likely to unlock new pathways for ultracold-atoms 
physics.

\appendix

\section{Floquet states of a periodically driven ``free'' particle}
\label{app:1}

In this Appendix~\ref{app:1} we construct the Floquet states of a 
periodically driven, but otherwise free particle moving in one spatial 
dimension, as described in the position representation by the Hamiltonian
\begin{equation}
	H(x,t) = -\frac{\hbar^2}{2M}\frac{\partial^2}{\partial x^2}
	- x F(t) \; ,
\label{eq:HAM}
\end{equation}	
where $M$ is the particle's mass. To this end, we adopt a strategy 
devised by Husimi for solving the Schr\"odinger equation
\begin{equation}
	\ri \hbar \frac{\partial}{\partial t} \psi(x,t)
	= H(x,t) \, \psi(x,t)
\label{eq:SGL}
\end{equation}
with a Hamiltonian that is at most quadratic in both momentum and 
position~\cite{Husimi53}, and do for the moment {\em not\/} assume that
the external force~$F(t)$ depends periodically on time~$t$. Following
Husimi, we introduce a classical trajectory $\xi(t)$ that will be specified
precisely later, and measure the position relative to that tajectory, writing
\begin{equation}
	y = x - \xi(t) \; .
\end{equation}  	
With $\psi(x,t)$ denoting the solution to Schr\"odinger's equation to be
determined, we define the wave function
\begin{equation}
	\widetilde\psi(y,t) = \psi(y +\xi(t), t) \; ,
\end{equation}
implying both
\begin{equation}
	\ri\hbar\frac{\partial}{\partial t}\psi =
	\ri\hbar\frac{\partial}{\partial t}\widetilde\psi -
	\ri\hbar\dot{\xi}\frac{\partial}{\partial y}\widetilde\psi
\end{equation}
and
\begin{equation}
	\frac{\partial}{\partial x} \psi =
	\frac{\partial}{\partial y} \widetilde\psi \; .	
\end{equation}
Hence, the dynamics of $\widetilde{\psi}$ are governed by
\begin{equation}
	\ri\hbar\frac{\partial}{\partial t} \widetilde{\psi} =
	\ri\hbar\dot{\xi}\frac{\partial}{\partial y} \widetilde{\psi} -
	\frac{\hbar^2}{2M} \frac{\partial^2}{\partial y^2} \widetilde{\psi} -
	F(t)(y + \xi)\widetilde{\psi} \; . 	
\end{equation}	
To account for the shift of momentum associated with the motion along $\xi(t)$,
we then perform a unitary transformation according to
\begin{equation}
	\widetilde{\psi}(y,t) = \exp\big(\ri M \dot{\xi}(t) y/\hbar\big) \,  
	\varphi(y,t) \; ,
\label{eq:UTR}
\end{equation}
so that the evolution of the wave function $\varphi(y,t)$ introduced in this
manner is given by
\begin{eqnarray}
	\ri\hbar\frac{\partial}{\partial t}\varphi & = & 
	-\frac{\hbar^2}{2M} \frac{\partial^2}{\partial y^2} \varphi 
	+ \Big( M \ddot{\xi} - F(t) \Big) y \varphi
\nonumber \\	& & 	
	- \Big( \frac{1}{2} M \dot{\xi}^2 + F(t) \xi \Big) \varphi \; . 
\label{eq:TEE}
\end{eqnarray}	
Let us now stipulate that the so far arbitrary trajectory $\xi(t)$ obeys
the classical equation of motion 
\begin{equation}
	M \ddot{\xi} = F(t) \; ,
\label{eq:CEM}
\end{equation}	
so that the second term on the right-hand side of Eq.~(\ref{eq:TEE}) vanishes.
The third term then is proportional to the classical Lagrangian
\begin{equation}
	L(t) = \frac{1}{2} M \dot{\xi}^2 + F(t) \xi
\label{eq:LAN}
\end{equation}
evaluated along this trajectory, so that Eq.~(\ref{eq:TEE}) takes the simple
form
\begin{equation}
	\ri\hbar\frac{\partial}{\partial t}\varphi = 
	-\frac{\hbar^2}{2M} \frac{\partial^2}{\partial y^2} \varphi 
	- L(t) \varphi \; . 
\label{eq:SIF}
\end{equation}
With the help of the ansatz
\begin{equation}
	\varphi(y,t) = \chi(y,t) \, \exp\!\left( \frac{\ri}{\hbar} 
	\int_0^t \! \rd \tau \, L(\tau) \right)
\label{eq:ANS}
\end{equation}
this reduces to the free-particle equation 
\begin{equation}
	\ri\hbar\frac{\partial}{\partial t}\chi = 
	-\frac{\hbar^2}{2M} \frac{\partial^2}{\partial y^2} \chi
\end{equation}
with the well-known plane-wave solutions
\begin{equation}
	\chi_k(y,t) = \exp\big[ \ri k y - \ri E(k) t/\hbar \big] 
\label{eq:PWS}
\end{equation}
labeled by a wave number~$k$, with energies
\begin{equation}
	E(k) = \frac{\hbar^2 k^2}{2M} \; .
\end{equation}	
Undoing the transformations~(\ref{eq:ANS}) and~(\ref{eq:UTR}), we thus have 
found improper solutions
\begin{eqnarray}
	\psi_k(x,t) & = & \exp\!\left( \ri k \big(x - \xi(t) \big) 
	+ \frac{\ri}{\hbar} M \dot{\xi(t)} \big(x - \xi(t) \big) \right)
\nonumber \\	& & \times
	\exp\!\left( \frac{\ri}{\hbar} \int_0^t \! \rd \tau \, L(\tau)
	 - \frac{\ri}{\hbar} E(k) t \right)
\label{eq:HWF}
\end{eqnarray}
to the original Schr\"odinger equation~(\ref{eq:SGL}) with the 
Hamiltonian~(\ref{eq:HAM}), from which one may easily construct proper, 
normalized wave packets with the help of suitable momentum weighting functions. 

While the above procedure applies to an arbitrary time-dependent force~$F(t)$,
we now require that $F(t)$ be $T$-periodic {\em without\/} a dc component, 
such as
\begin{equation}
	F(t) = F_1 \cos(\omega t)
\label{eq:SSF}
\end{equation}	
with $T = 2\pi/\omega$. Then the classical equation of motion~(\ref{eq:CEM})
possesses a particular $T$-periodic solution,
\begin{equation}
	\xi(t) = -\frac{F_1}{M\omega^2} \cos(\omega t) \; , 
\label{eq:CLT}
\end{equation}	
giving the $T$-periodic Lagrangian
\begin{equation}
	L(t) = \frac{F_1^2}{2M\omega^2} \sin^2(\omega t) 
	- \frac{F_1^2}{M\omega^2} \cos^2(\omega t) \; .
\end{equation}
However, the time-integrated Lagrangian appearing in the wave 
function~(\ref{eq:HWF}) generates a secular term that grows linearly in time, 
with the growth rate being determined by the zero mode
\begin{equation}
	\frac{1}{T} \int_0^T \! \rd \tau \, L(\tau) 
	= -\frac{F_1^2}{4M\omega^2} \; . 
\label{eq:ZEM}
\end{equation} 
For writing the solutions~(\ref{eq:HWF}) in the desired Floquet form
\begin{equation}
	\psi_k(x,t) = u_{k,m}(x,t) \exp(-\ri\varepsilon_{k,m} t/\hbar) 
\end{equation}
with duly $T$-periodic Floquet functions $u_{k,m}(x,t) = u_{k,m}(x,t+T)$,
one only has to realize that this zero mode contributes to the quasienergies 
$\varepsilon_{k,m}$, resulting in
\begin{eqnarray}
	& &
	u_{k,m}(x,t) = \exp\!\left( \ri k \big(x - \xi(t) \big) 
	+ \frac{\ri}{\hbar} M \dot{\xi(t)} \big(x - \xi(t) \big) \right)
\nonumber \\	& \times &
	\exp\!\left( \frac{\ri}{\hbar} \left[ \int_0^t \! \rd \tau \, L(\tau)
	- \frac{t}{T} \! \int_0^T \! \rd \tau \, L(\tau) \right] 
	+ \ri m \omega t \right)
\end{eqnarray}
and
\begin{eqnarray}
	\varepsilon_{k,m} & = & E(k) + m \hbar\omega 
	- \frac{1}{T} \! \int_0^T \! \rd \tau \, L(\tau) 
\nonumber \\	& = &
	\frac{\hbar^2 k^2}{2M} + m \hbar\omega + \frac{F_1^2}{4M\omega^2} \; . 
\label{eq:QEF}
\end{eqnarray}	
Here the integer~$m$ designates the respective Floquet representative, 
as in Eqs.~(\ref{eq:REP}) and (\ref{eq:FUM}). The third contribution to 
$\varepsilon_{k,m}$, that is, the negative of the zero mode~(\ref{eq:ZEM}), 
is just the average quiver energy of a particle moving under the influence 
of the sinusoidal force~(\ref{eq:SSF}), analogous to the ``ponderomotive 
energy'' considered in the study of matter interacting with strong laser 
fields~\cite{KruchininEtAl18,Friedrich17,DellaPiccaEtAl16}. Note that the 
argument employed here for extracting the quasienergies~(\ref{eq:QEF}) from 
the Husimi solutions~(\ref{eq:HWF}) precisely parallels the one invoked in 
Sec.~\ref{sec:3} for extracting the quasienergy band~(\ref{eq:QEB}) from the 
Houston functions~(\ref{eq:HOU}). Observe also that the Floquet states derived 
from the solutions~(\ref{eq:HWF}) are closely related to the well-known 
nonrelativistic Gordon-Volkov wave functions~\cite{Friedrich17,
DellaPiccaEtAl16,Gordon26,Wolkow35}. 

Next, let us assume that the external force {\em does\/} contain a
dc contribution,
\begin{equation}
	F(t) = F_0 + F_1 \cos(\omega t)
\label{eq:FUF}
\end{equation}
with nonzero $F_0$. In this case the previous procedure again allows one 
to construct Husimi solutions~(\ref{eq:HWF}) to Schr\"odinger's equation,
but it is no longer possible to obtain Floquet states from these wave
functions, since there are no $T$-periodic solutions to the classical equation
of motion~(\ref{eq:CEM}). Therefore, we proceed in a slightly different manner: 
If we again select the $T$-periodic trajectory~(\ref{eq:CLT}) that embodies 
only the ac component of the forcing, we obtain the equation
\begin{equation}
	\ri\hbar\frac{\partial}{\partial t}\varphi = 
	-\frac{\hbar^2}{2M} \frac{\partial^2}{\partial y^2} \varphi 
	- F_0 y \varphi - L(t) \varphi 
\end{equation}
instead of Eq.~(\ref{eq:SIF}), where now $L(t)$ is the 
Lagran\-gian~(\ref{eq:LAN}) with the full force~(\ref{eq:FUF}), but still 
being evaluated along the trajectory~(\ref{eq:CLT}). This guarantees that 
$L(t)$ again is $T$-periodic, containing the same zero mode~(\ref{eq:ZEM}) 
as before. The transformation~(\ref{eq:ANS}) then leads to the equation    
\begin{equation}
	\ri\hbar\frac{\partial}{\partial t}\chi = 
	-\frac{\hbar^2}{2M} \frac{\partial^2}{\partial y^2} \chi
	-F_0 y \chi \; ,
\end{equation}
the admissible solutions of which are given by 
\begin{equation}
	\chi_E(y,t) = {\rm Ai}\left(\frac{y_0 - y}{a}\right)
	\exp(-\ri Et/\hbar) \; , 
\label{eq:AFU}
\end{equation}	
where ${\rm Ai}(z)$ is an Airy function~\cite{AbramowitzStegun72},
\begin{equation}
	a = \left(\frac{\hbar^2}{2MF_0}\right)^{1/3} \; ,
\end{equation}
and
\begin{equation}
	y_0 = -E/F_0 
\end{equation}
for arbitrary energy~$E$. These functions~(\ref{eq:AFU}) now play the role 
of the previous plane-wave solutions~(\ref{eq:PWS}); as before, the Floquet 
states corresponding to the ac-dc forcing~(\ref{eq:FUF}) are obtained by 
reverting the transformations~(\ref{eq:ANS}) and~(\ref{eq:UTR}), and by 
extracting the zero mode of the time-integrated Lagrangian. In particular, 
this implies that the quasienergies of a ``free'' particle subjected to the 
force~(\ref{eq:FUF}) are given by
\begin{equation}
	\varepsilon_{E,m} = E + m\hbar\omega + \frac{F_1^2}{4M\omega^2} \; .
\label{eq:PES}
\end{equation}	  
Thus, in both cases~(\ref{eq:SSF}) and (\ref{eq:FUF}) the dependence of the
quasienergies on the amplitude $F_1$ of the driving force is given by the 
ponderomotive energy shift, which explains the ``background'' observed in
Fig.~\ref{F_2}.

\section{Essentials of the box stabilization method}
\label{app:2}

In order to provide an illustrative example for the use of the so-called box 
stabilization method for computing resonance energies~$E^{\rm res}$ and total 
widths~$\Gamma$ of metastable decaying states~\cite{MandelshtamEtAl93,
MakarovEtAl94} we consider the proto\-typical model potential 
\begin{equation}
	V(x) = \left\{
	\begin{array}{lrl} 
		\infty 	\quad , &   		& x \leq 0	\\
		0 \;\;	\quad , &	0 <  	& x < a 	\\
		V_0	\quad , &	a \leq	& x \leq a+b	\\
		0 \;\;	\quad , &	a+b < 	& x \; .
	\end{array} \right. 	 	
\label{eq:MOP}
\end{equation}
The region between the hard wall located at $x = 0$ and the barrier extending 
from $x = a$ to $x = a+b$ with positive height~$V_0$ may support long-lived, 
metastable states that ``decay'' by tunneling. While the energy spectrum of 
the Hamiltonian
\begin{equation}
	H(x) = -\frac{\hbar^2}{2M} \frac{\rd^2}{\rd x^2} + V(x)
\label{eq:MOH}
\end{equation}
is continuous, admitting improper, non-normalizable eigenstates only, such a 
``resonance'' manifests itself through a Lorentzian peak in the density of 
states~\cite{Titchmarsh51}, 	    
\begin{equation}
	\rho(E) \approx \frac{1}{\pi} 
	\frac{\Gamma/2}{(E - E^{\rm res})^2 + (\Gamma/2)^2} \; ,
\label{eq:LOK}
\end{equation}	
with $\tau = \hbar/\Gamma$ being the $1/{\rm e}$ lifetime of the decaying 
resonant state. A well-established method for determining the resonance 
parameters $E^{\rm res}$ and $\Gamma$ relies on a complex rotation 
\begin{equation}
	x \to x\re^{\ri\theta}
\label{eq:DIT}
\end{equation}
of the coordinate~$x$ by an angle $\theta$, that is, on a dilatation 
transformation~\cite{AguilarCombes71,BalslevCombes71,Simon73}. Thus, the
plane waves $\re^{\pm\ri kx}$ that constitute the eigenfunctions of the
Hamiltonian~(\ref{eq:MOH}) for $x > a+b$ become exponentially increasing
or decreasing functions:
\begin{equation}
	\re^{\pm\ri kx} \to \re^{\pm\ri k x \cos(\theta)}
	\re^{\mp k x \sin(\theta)} \; . 
\end{equation}	
Therefore, the non-unitary transformation~(\ref{eq:DIT}) convertes the
original Hermitian Hamiltonian into a non-Hermitian one that possesses
square-integrable eigenstates with complex energies
\begin{equation}
	E = E^{\rm res} - \ri \frac{\Gamma}{2} \; .
\end{equation}
In the case of the model~(\ref{eq:MOP}), an elementary calculation shows 
that these complex energy eigenvalues are given by the solutions to the 
equation 
\begin{equation}
	-\ri = \frac{\sin(ka) \cosh(\kappa b) 
		+ (k/\kappa) \cos(ka)\sinh(\kappa b)}
		    {\cos(ka) \cosh(\kappa b)
		+ (\kappa/k) \sin(ka)\sinh(\kappa b)} \; ,  
\label{eq:TRA}
\end{equation}
where
\begin{equation}
	k = \frac{1}{\hbar}\sqrt{2ME}
\end{equation}
and
\begin{equation}
	\kappa = \frac{1}{\hbar} \sqrt{2M(V_0 - E)} \; .
\end{equation}

\begin{figure}[t]
\begin{center}
\includegraphics[width = 0.9\linewidth]{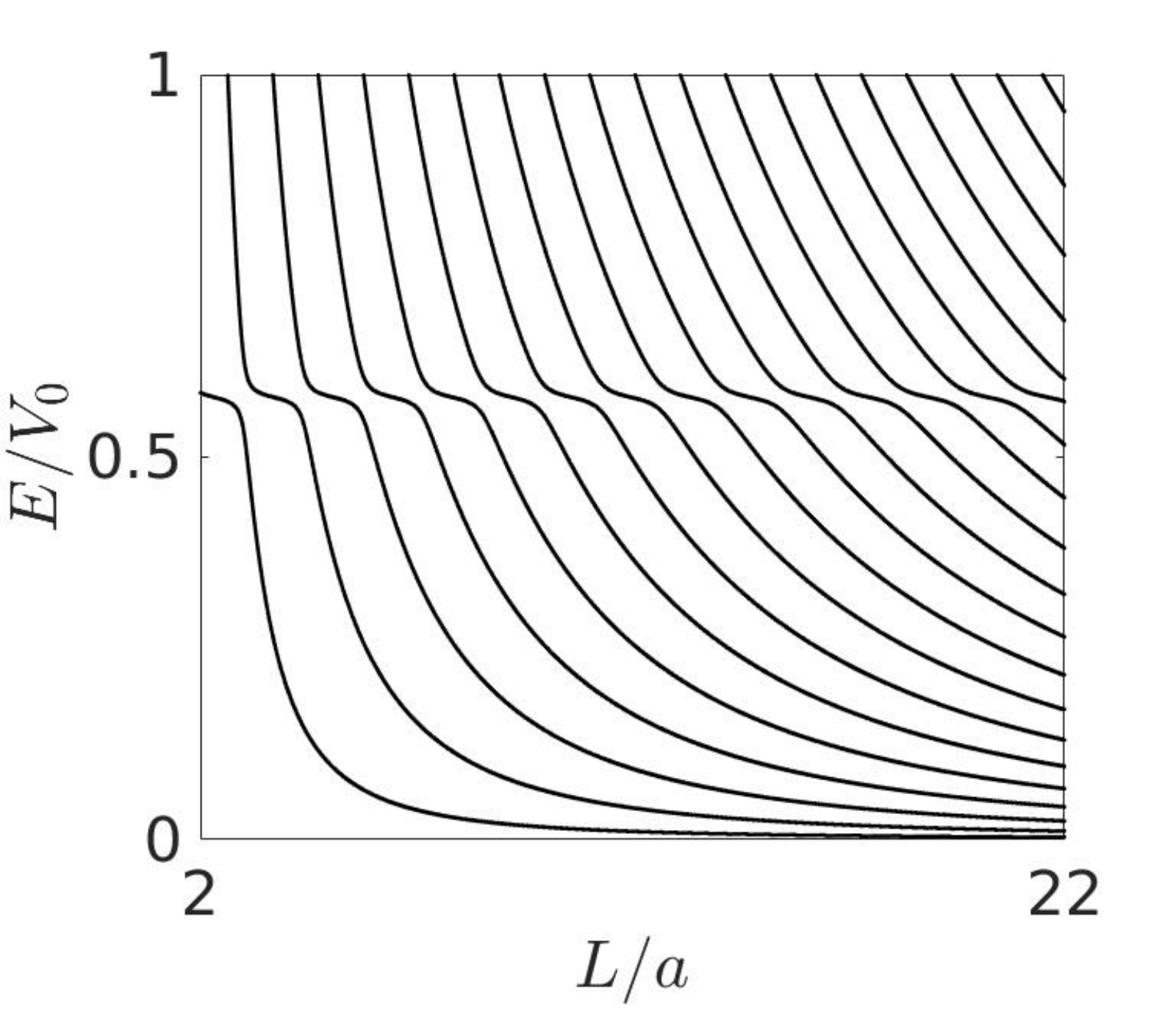}
\end{center}
\caption{Stabilization diagram for the model potential~(\ref{eq:MOP}) 
	with barrier width $b/a = 1$ and height $2Ma^2V_0/\hbar^2 = 9$. 
	Shown are the eigenvalues $E_n$ of the boxed system $H_L(x)$ vs.\ 
	the box size~$L$. The precursor of the resonance is visible clearly; 
	the widths of the avoided eigenvalue crossings translate into its 
	decay width~$\Gamma$.}      
\label{F_B1}
\end{figure}

The stabilization method is based on the recognition that information 
on the decaying system can be obtained by confining it to a box, that is, 
by erecting a hard wall at some position $L > a+b$, outside the range of 
the potential. Due to the resulting Dirichlet boundary condition at 
$x = L$, the modified Hamiltonian $H_L(x)$ posseses only a countable set of 
square-integrable eigenstates; each of these could be continued to the regime 
$x > L$ such that it coincides with a non-normalizable eigenstate of the 
original $H(x)$. In Fig.~\ref{F_B1} we depict the corresponding stabilization 
diagram, displaying the eigenvalues of the ``boxed'' system $H_L(x)$ vs.\ the 
box size~$L$. If the barrier were impenetrable, the resonant state localized 
to its left would not be affected at all by the variation of $L$; due to the 
nonzero tunnel coupling, its energy actually exhibits an avoided crossing 
whenever hitting one of the energies associated with continuum-like box states 
living to the right of the barrier.     

For each box size~$L$, the density $\rho_L(E)$ of eigenstates of $H_L(x)$ 
is given by a sequence of delta spikes situated at its energy eigenvalues 
$E_n(L)$,
\begin{equation}
	\rho_L(E) = \sum_n \delta\big( E - E_n(L) \big) \; .
\label{eq:DDS}
\end{equation}	
The key proposition of the box stabilization approach is that a good 
approximation to the continuous spectral density $\rho(E)$ pertaining 
to the original, unboxed system $H(x)$ is obtained by averaging these 
densities~(\ref{eq:DDS}) over a suitable interval of the auxiliary 
parameter~$L$:
\begin{equation}
	\rho(E) \approx \frac{1}{\Delta L} \int_{\Delta L} \!\! \rd L \;
	\rho_L(E) \; .
\end{equation}
Integrating over a small energy bin $\Delta E$, one then also has
\begin{eqnarray}
	\rho(E) \Delta E & \approx & \frac{1}{\Delta L} 
	\int_{\Delta E} \!\! \rd E \int_{\Delta L} \!\! \rd L \; \rho_L(E)
\\	& = &
	\frac{1}{\Delta L} \int_{\Delta L} \! \rd L \int_{\Delta E} \! \rd E
	\, \sum_n \delta\big( E - E_n(L) \big) \; .				
\nonumber
\end{eqnarray}
If now the integral over~$L$ is replaced by a sum over discrete values $L_i$ 
evenly spaced by $\delta L$, one finally arrives at
\begin{equation}
	\rho(E) \Delta E \approx \frac{\delta L}{\Delta L}
	\sum_i \sum_n \int_{\Delta E} \!\! \rd E \; 
	\delta\big( E - E_n(L_i) \big) \; .	
\label{eq:HIS}
\end{equation}
This translates into a convenient numerical recipe: Select an interval 
$\Delta L$ ranging from some $L_{\min} \geq a+b$ to $L_{\max}$, discretize 
this interval with evenly spaced mesh points $L_i$, and compute the energy
eigenvalues $E_n(L_i)$ for each of these box sizes. Then count the total
number of eigenvalues falling into an energy bin $\Delta E$; apart from the
factor $\delta L/\Delta L$ the resulting sum is just the right hand side
of Eq.~(\ref{eq:HIS}). Thus, a sufficently fine binning of the energy axis
allows one to construct the desired spectral density $\rho(E)$.     

The computational power of this approach derives from the fact that it is 
{\em not} necessary to employ large box sizes for approximating a continuum;
expensive diagonalizations with large basis sets as would be required for 
large~$L$ give way to a sequence of comparatively cheap diagonalizations for 
small boxes with correspondingly small basis sets. But the decisive insight 
to be gained in the present context stems from the observation that it is 
the width of the avoided eigenvalue crossings appearing in the stabilization 
diagram that translates directly into the deacy width~$\Gamma$, thereby 
determining the lifetime $\tau = \hbar/\Gamma$ of the resonance. It is this 
very feature that is exploited when adapting the box stabilization method 
to the computation of densities of Floquet-Bloch states, and to the estimation
of escape times in Sec.~\ref{sec:4}.

\end{document}